\documentclass{ar-1col-S2O}

\usepackage[comma]{natbib}
\usepackage{url}
\usepackage{amssymb}
\usepackage{graphicx}
\usepackage{subfig}
\usepackage{soul}

\doi{10.1146/((please add article doi))}

\newcommand{\micron}{\mbox{$\mu$m}}
\newcommand{\arcsec}{\mbox{$^{\prime\prime}$}}
\newcommand{\arcmin}{\mbox{$^{\prime}$}}

\newcommand{\nthp}{\mbox{N$_2$H$^+$}}

\newcommand{\nht}{\mbox{NH$_3$}}
\newcommand{\hcop}{\mbox{HCO$^+$}}

\newcommand{\cateo}{\mbox{C$^{18}$O}}
\newcommand{\csevo}{\mbox{C$^{17}$O}}
\newcommand{\thco}{\mbox{$^{13}$CO}}
\newcommand{\twco}{\mbox{$^{12}$CO}}

\newcommand{\lsun}{\mbox{L$_{\odot}$}}
\newcommand{\msun}{\mbox{M$_{\odot}$}}
\newcommand{\mstar}{\mbox{M$_{\rm *}$}}

\newcommand{\tbol}{\mbox{T$_{\rm bol}$}}
\newcommand{\lbol}{\mbox{L$_{\rm bol}$}}
\newcommand{\menv}{\mbox{M$_{\rm env}$}}
\newcommand{\farcs}{\mbox{.$^{\prime\prime}$}}

\begin{document}

\markboth{Tobin \& Sheehan et al.}{Protostars}

\title{An Observational View of Structure in Protostellar Systems}

\author{John J. Tobin,$^1$ and Patrick D. Sheehan,$^{1}$
\affil{$^1$National Radio Astronomy Observatory, Charlottesville, VA, USA, 22903; email: jtobin@nrao.edu}
}

\begin{abstract}
The envelopes and disks that surround protostars reflect the initial conditions of star and planet formation and govern the assembly of stellar masses. Characterizing these structures requires observations that span the near-infrared to centimeter wavelengths. Consequently, the past two decades have seen progress driven by numerous advances in observational facilities across this spectrum, including the \textit{Spitzer Space Telescope}, \textit{Herschel Space Observatory}, the Atacama Large Millimeter/submillimeter Array, and a host of other ground-based interferometers and single-dish radio telescopes. 
Nearly all protostars appear to have well-formed circumstellar disks that are likely to be rotationally-supported; the ability to detect a disk around a protostar is more a question of spatial resolution than whether or not a disk is present. The disks around protostars have inherently higher millimeter/submillimeter luminosities as compared to disks around more-evolved pre-main sequence stars, though there may be systematic variations between star forming regions. The envelopes around protostars are inherently asymmetric and streamers emphasize that mass flow through the envelopes to the disks may not be homogeneous. The current mass distribution of protostars may be impacted by selection bias given that it is skewed toward solar-mass protostars, inconsistent with the stellar initial mass function.

\end{abstract}

\maketitle
\tableofcontents

\section{Introduction}
The material surrounding newborn stars plays a pivotal role in the lives of stars and planets as the fuel of their formation. The morphology of the gas and dust is multi-scale, with different physical structures being the dominant source of emission (or mass) on different spatial scales, and this material needs to overcome supporting forces and collapse or fall-in to form the protostar. The infalling material must also conserve angular momentum, which results in the formation of a disk around the protostar that ultimately plays a key role in regulating the mass accretion onto the protostar. These disks that form in the protostellar phase are the progenitors to protoplanetary disks \citep[see review by][]{Andrews2020}, where planets have long been thought to form, but more recent work has suggested that the disks around protostars are very likely the place where the planet formation process begins.

A protostellar system is composed of intertwined structures where the infalling envelope influences the disk, and the disk influences the mass accretion rate onto the protostar, and the protostellar mass should influence the infall and accretion rates. We will focus primarily on low-mass protostars in this review, while high-mass star formation is covered in a recent review \citep{Motte2018}. The interaction of the accretion disk and the protostar drives bipolar outflows \citep[e.g.,][]{Bally2016} that excavate cavities in the polar regions of envelopes, representing an important feedback process and creating a low-extinction pathway for near to mid-infrared light to escape from the protostar and disk. The amount of material within the infalling envelope will provide a limit on the amount of mass that a protostar may accrete during the time it takes for the envelope to fall-in completely from a characteristic radius. Envelopes, however, are connected to their surrounding environments \citep[e.g.,][]{Pineda2020}, so the instantaneous envelope mass may not provide an ultimate limit on the amount of mass a protostar may accrete. The disk then regulates the rate at which material flows through it and onto the protostar, but also whether companions (or giant planets) may form via gravitational instability, and also will ultimately evolve into a protoplanetary system. 

\begin{figure}
\begin{center}
\includegraphics[scale=0.5]{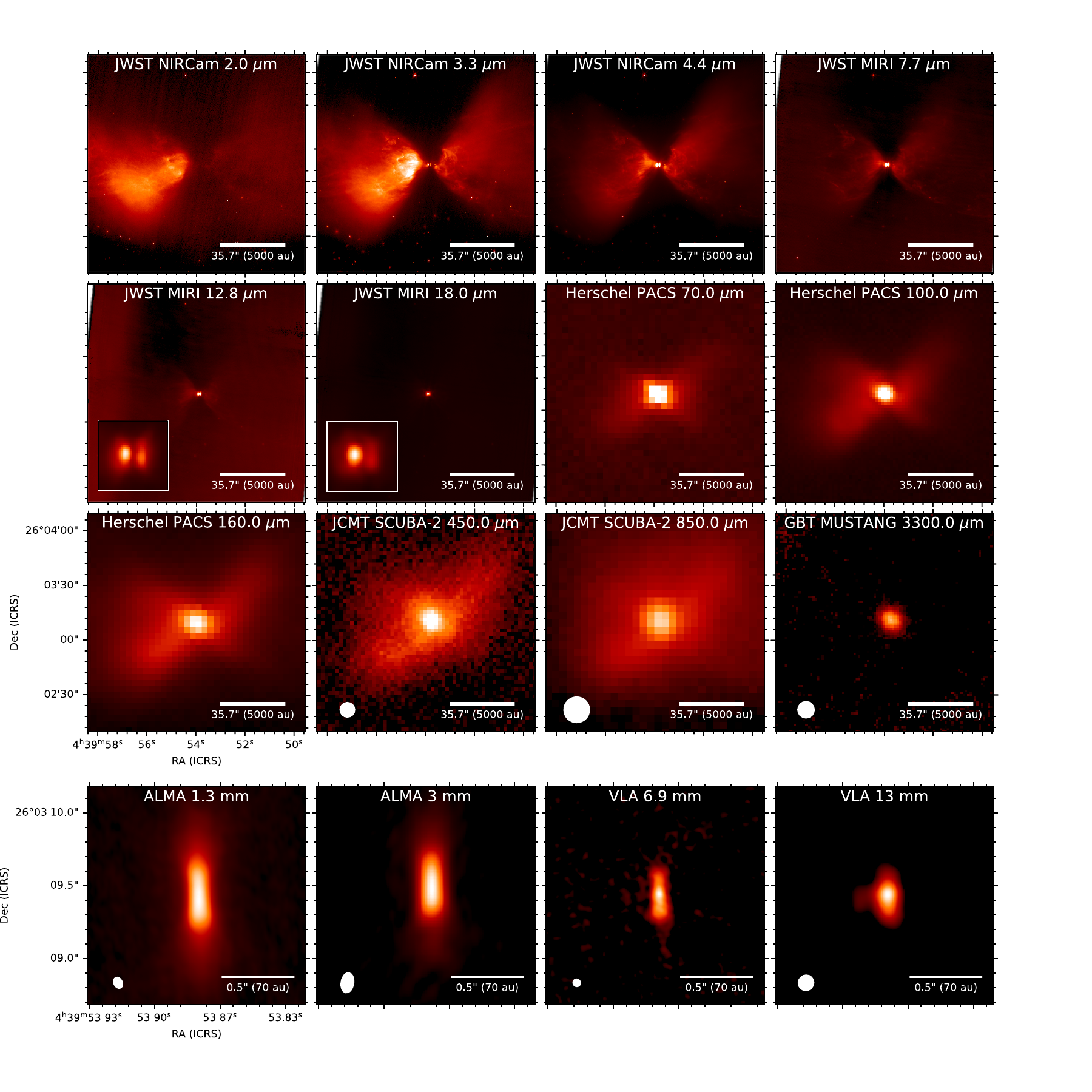}
\end{center}
\vspace{-5mm}
\caption{Multi-wavelength images of the protostar L1527 IRS from 2~\micron\ to 13~mm plotted using a square-root color stretch. The top three rows show emission on 2\arcmin\ scales from the highest resolution data available, and the bottom row shows the highest resolution data available from millimeter to centimeter-wave interferometers, focusing on the small-scale disk emission. The insets for the 12.8 and 18~\micron\ data are 4\arcsec\ (560 au) wide, showing that the dark lane from the edge-on disk persists into the mid-infrared. The \textit{James Webb Space Telescope} (JWST) NIRCam and MIRI images are from the JWST archive (Program ID: 2739), the \textit{Herschel} data were observed as part of the Gould Belt Survey \citep{Andre2010}, the SCUBA-2 data are from the James Clerk Maxwell Telescope Archive, the GBT Mustang data are from \citet{Shirley2011}, the ALMA 1.3~mm data are from \citep{vanthoff2023}, the ALMA 3~mm data are from \citet{Nakatani2020}, and the VLA data are from \citep{Sheehan2022b}.}
\label{L1527_multiwave}
\end{figure}

\begin{figure}
\includegraphics[scale=0.75]{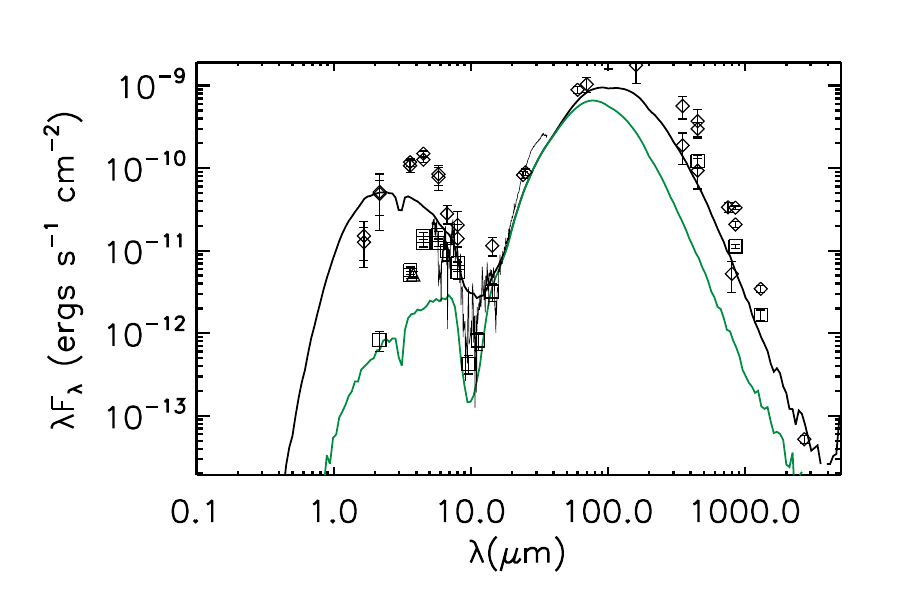}
\vspace{5mm}
\caption{SED of the protostar L1527 IRS, reproduced from \citet{Tobin2010TheImaging}. The photometry are shown for two apertures, one 71\farcs4 (10000 au) in radius (diamonds) and the other 7\farcs14 (1000 au) in radius (boxes). The continuous lines are model SEDs measured in 10000 au (black) and 1000 au (gray) radius apertures. The \textit{Spitzer} IRS spectrum is also overlaid from 5 to 35~\micron. The overlap in the model SED lines between $\sim$10~\micron\ to $\sim$60~\micron\ further demonstrates that the emission at these wavelengths is dominated by small radii in the inner envelope and disk, similar to what is shown spatially in Figure \ref{L1527_multiwave}.}
\label{L1527_SED}
\end{figure}

To reveal the nature of protostellar systems to the fullest possible extent, they need to be studied with a multi-wavelength point of view. The readily detectable emission from a protostar system generally spans $\sim$1~\micron\ to $\sim$10~cm, but there are a few cases where protostars exhibit X-ray emission \citep{Grosso2020}. We show multi-wavelength and multi-scale images of the protostar L1527 IRS in Figure \ref{L1527_multiwave}, demonstrating the continuum emission processes and their scales within protostellar systems. Light from the protostar and inner disk scattering on dust grains within the outflow cavities and cavity walls dominates the $\sim$1-10~\micron\ emission, while from 10 to $\sim$100~\micron\ thermal dust emission from warm dust close to the protostar dominates the emission, and then $\sim$100~\micron\ to 3~mm is also dominated by thermal dust emission, but colder dust farther from the protostar. The higher resolution enabled by interferometers probes the thermal dust emission from the disk surrounding the protostar, while being much less sensitive to the envelope emission. Emission at wavelengths beyond $\lambda$~$\sim$~1~cm (the longest shown in Figure \ref{L1527_multiwave} being 1.3~cm) is dominated by thermal free-free emission, and in some cases non-thermal synchrotron emission for low-mass protostars, both of which are powered by the jets and outflows from the protostar \citep{Tychoniec2018}. The spectral energy distribution (SED) of the protostar L1527 IRS is shown in Figure \ref{L1527_SED} within different sized apertures, further demonstrating the scales from which the emission is emitted as a function of wavelength.

Studying the dynamics of star formation (e.g., protostellar collapse, disk formation, outflows, etc.) require the observations of spectral lines emitted by molecules or atoms tracing the structures of interest. The motion of the gas along the line of sight is traced via Doppler-shifted molecular line emission. This review does not cover astrochemistry \citep[see review by][]{Jorgensen2020}, but we do discuss molecular lines that typically trace the dynamics of protostar systems. The cold (T~$<$~100~K) regions in protostellar systems are typically traced by submillimeter/millimeter line emission, while warmer regions (T~$>$~100~K) are typically traced with mid-to-far-infrared line emission.

The necessity of multi-wavelength observations to characterize protostellar systems also implies that advances in their study will follow the advancement of observational facilities. The studies of protostars have evolved greatly in the past two decades and each new instrument, ground or space-based, has revealed previously unknown or unappreciated physics of the star and planet formation process given the increase in sensitivity, resolution, and/or survey speed. The rapid pace of advancement is demonstrated by the fact most of the images in Figure \ref{L1527_multiwave} are less than a decade old, and the advancements of the past 15 years have provided a completely refreshed view of protostellar systems. 
In this review, we will explore the advances of this rich period of discovery for protostellar systems. 

\subsection{The Anatomy of a Protostellar System}

The components considered to make up a protostellar system are a product of the research spanning more than four decades.
However, many terms can be used in different ways throughout the literature and usage may differ between primarily observation work and theoretical/numerical studies. Thus, it is appropriate to establish the basic definitions that we will use to describe the different components of a protostellar system.

\subsubsection{Cloud}
We refer to a \textit{cloud} (or a molecular cloud) as a lower density (n $<$ 10$^{5}$~cm$^{-3}$) medium within which the protostellar core and other structures are embedded. This gas is not bound to a protostar.

\subsubsection{Filament}
The molecular clouds themselves are found to be threaded with filaments \citep{Andre2014} and starless, pre-stellar, and protostellar cores within molecular clouds are predominantly found within filamentary structures. Like the molecular clouds themselves, these structures are not bound to the protostar.

\subsubsection{Core}
A core is the a concentration of dense gas (n $\gtrsim$ 10$^{4}$~cm$^{-3}$) that is expected to be 1000s of au to $\sim$0.1~pc across, and they can have a variety of morphologies/shapes. The term core does not uniquely refer to protostellar systems and is often used to describe concentrations of gas or dust that could be starless, pre-stellar, or protostellar. A starless core simply means there is no protostar present and it may or may not go on to form a protostar, while a pre-stellar core has had its conditions characterized and it is expected to collapse and form a protostar. Core sometimes can also be used to refer to the protostar itself (see section 1.1.10), leading to some confusion between theoretical and observational literature.

\subsubsection{Envelope}
The envelope and core are not mutually exclusive, but the envelope specifically refers to the infalling region of the core, which could encompass the entire core or only part of the core. Envelopes and cores are often also given defined boundaries, but aside from some of the truly isolated cores \citep[e.g., B335, ][]{Stutz2008}, the cores and envelopes are connected to a larger scale molecular cloud.

\subsubsection{Streamer}
A streamer is a thin (high-aspect ratio) portion of the envelope or core that is falling-in toward the central protostar and disk. Streamers often appear as enhanced molecular line emission and can be both large- ($\sim$10000~au) and small-scale ($\sim$1000~au) and may have enhanced density. Streamers could be related to the larger-scale filaments and may highlight the infall of asymmetric material both within the envelope and from the molecular cloud to the envelope.

\subsubsection{Outflow cavity}
 Outflow cavities are generally the polar regions within the protostellar envelope (orthogonal to the disk plane) where the outflow has removed envelope material. These cavities are generally conical in nature and provide a low extinction pathway for light from the protostar and inner disk to escape, which then scatters on dust grains making them apparent in the near to mid-infrared. 
 
\subsubsection{Disk}
The disk is a rotationally-supported structure that forms inside the envelope as a consequence of angular momentum conservation. The terms `embedded disk' or `protostellar disk' are often used to distinguish between disks around young protostars still embedded within their natal envelopes and disks around pre-main sequence stars with little envelope material remaining, which are typically referred to as `protoplanetary' disks.

\subsubsection{Pseudo-disk}
This is a confusing term that specifically refers to a magnetically-dominated, flattened, infalling structure that would correspond to an envelope. The flattened nature leads to confusion of pseudo-disks with disks for which Keplerian rotation had not yet been identified. The use of this term is most appropriate when there is a preponderance of evidence that a magnetically-dominated infalling structure is the best description of the observed structure, and there may only be one likely candidate thus far \citep[e.g.,][]{Maury2018}. We avoid this term in this review.

\subsubsection{First Hydrostatic Core}
Often abbreviated as FHSC, this is a theorized progenitor to the protostar, but the temperature is not high enough to dissociate molecular hydrogen. Once the temperature exceeds $\sim$2000~K, molecular hydrogen dissociates, changing the gas equation of state, causing it to contract and form the second hydrostatic core. 

\subsubsection{Protostar}
The protostar is the forming stellar object supported by hydrostatic equilibrium. The protostar is not yet undergoing nuclear fusion in its core, but has a high enough temperature that molecular hydrogen is dissociated. This is sometimes referred to as the second hydrostatic core.

\subsection{Overview of Protostar Classification}
\label{section:classes}

\begin{marginnote}[]
\entry{Young Stellar Object (YSO)}{A catch-all term to describe systems from protostars to pre-main-sequence stars.}
\end{marginnote}
The primary classification systems for young stellar objects (YSOs) have been in use with little refinement over the past three decades. We review these methods in order to address their strengths, weaknesses, and, as we will discuss in later sections, paths for improvement. It is important to keep in mind that protostars are classified by the light they emit, requiring observations from near-infrared to millimeter wavelengths, as demonstrated visually in Figures \ref{L1527_multiwave} and \ref{L1527_SED}.

YSOs have traditionally been classified according the the slope of their near-infrared spectral index 
\begin{equation}
\alpha=\frac{d \log (\lambda F_\lambda)}{d \log \lambda} \qquad.
\end{equation}
They were initially grouped into three categories, Class I with a rising near-infrared spectrum  ($\alpha > 0$), Class II with a shallow decline in the near-infrared ($0 < \alpha < -3$), and Class III with a rapidly declining near-infrared spectrum ($\alpha < -3$) approaching a stellar photosphere with a small amount of excess \citep{Lada1987}. Physically this was thought to trace a dusty infalling envelope enshrouding the protostar (Class I), a disk producing an infrared excess above the stellar photosphere (Class II), and ultimately a meager disk producing a small amount of excess (Class III). Flat Spectrum sources, with $-0.3 < \alpha < 0.3$, were added later \citep{Greene1994}, and were hypothesized as a transition phase between Class I and Class II.

A major addition to the system are the Class 0 sources \citep{Andre1993}: protostars so heavily embedded that they were not readily detectable in the near-infrared observations of the time, but very luminous at submillimeter/millimeter. The proposed criteria for Class 0 was L$_{\rm submm}$/L$_{\rm bol}$ $>$ 0.005. Bolometric luminosity in the context of this review is defined as
\begin{equation}
\rm L_{\rm bol} = 4\pi d^2\int_{0}^{\infty} F_{\lambda} d\lambda, 
\end{equation}
where $F_{\lambda}$ is the flux density at a given wavelength and $d$ is the distance to the protostar. L$_{\rm bol}$ is purely an empirical quantity measured by integration of the observed SED, without correction for the extincted optical and UV emission because most of this emission should be absorbed and reemitted by the envelope and disk. Also, for practical purposes, the starting and ending wavelengths for the measurement of L$_{\rm bol}$ are $\sim$1~\micron\ and $\sim$3~mm, respectively (see Figure \ref{L1527_SED}). Then to obtain L$_{\rm submm}$, one begins the integral at 350~\micron. The coverage of the SED is also not continuous and numerical integration, like the trapezoidal rule, is typically used. The relationship between L$_{\rm bol}$ and L$_{\rm total}$ can be complex, depending on the inclination of the protostar and how opaque the envelope is. Modeling of the SED using radiative transfer can provide estimates for L$_{\rm total}$ \citep[e.g.,][]{Whitney2003}.

\begin{figure}
\includegraphics[scale=0.15]{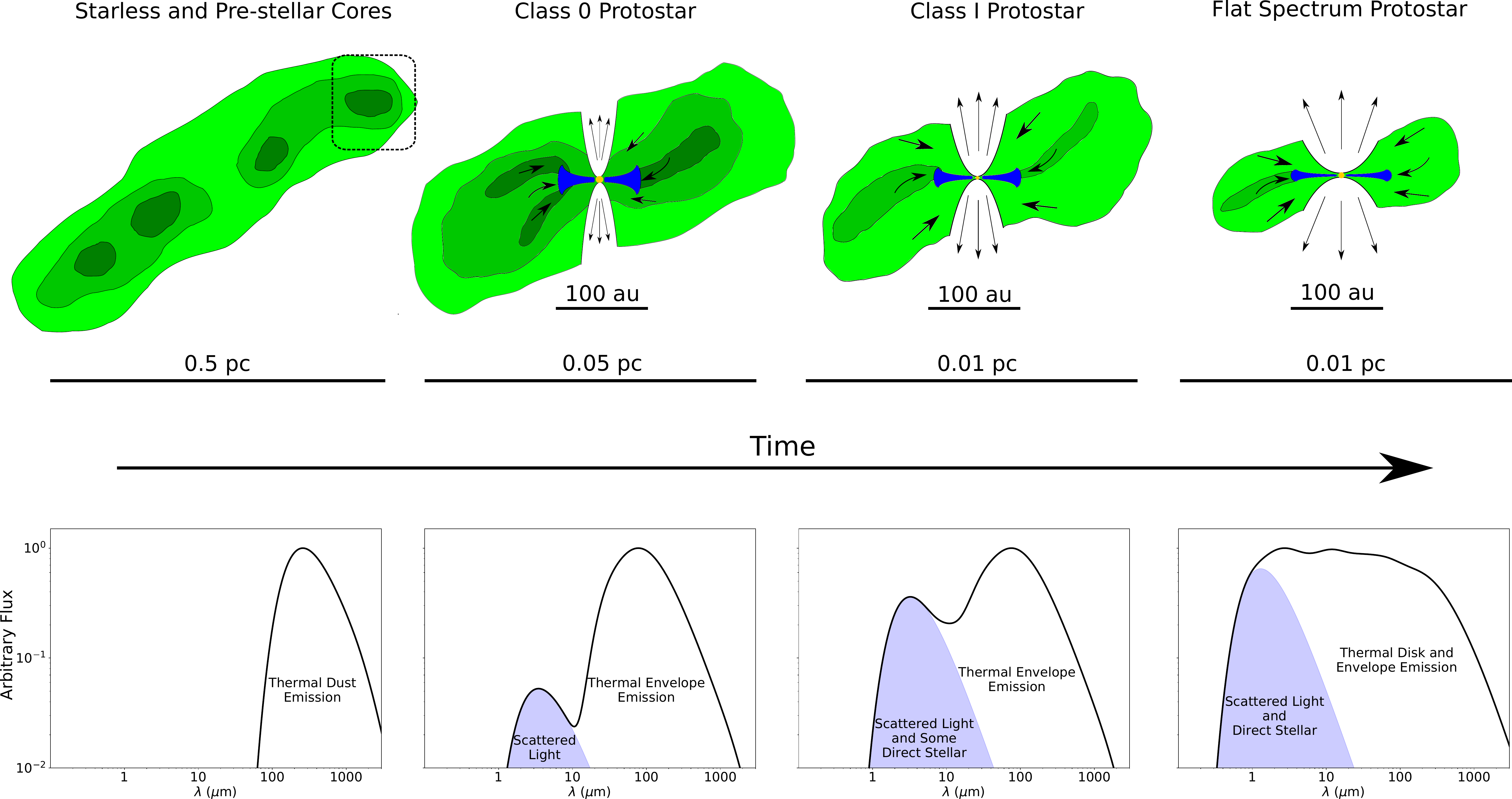}
\vspace{5mm}
\caption{Illustration of the star formation process for a single protostar with accompanying SEDs. The formation of a star begins with the collapse of a dense core that may be within a filament that is part of a larger-scale molecular cloud. The dashed box shows the area around a single core that is being zoomed-in on for the later classes. The collapsing core forming a Class 0 protostar has an asymmetric morphology, possibly filamentary, reflecting the structure of the progenitor starless core. As the envelope is accreted onto the disk and protostar, the density of the envelope is reduced and the protostar and disk are better visible through the envelope in the Class I phase. Then, as the density is further reduced and the outflow widens, the Class I protostar evolves into a Flat Spectrum protostar where the disk and protostar begin to dominate the SED rather than the envelope. Throughout the Class 0, I, and Flat Spectrum phases there could be dense channels of matter flowing onto the disk, which are typically called streamers (see Section 2.4).}
\label{overview}
\end{figure}

A complementary classification method is bolometric temperature \citep[\tbol;][]{Myers1993,Chen1995}, which is computed to be the temperature of a blackbody with the same flux-weighted
mean frequency as the SED. \citet{Chen1995} proposed the divisions between Classes in \tbol\ to be 
\begin{itemize}
\item Class 0:  70~K~$>$~\tbol~$>$~20~K
\item Class I: 650~K~$>$~\tbol~$>$~70~K
\item Class II: 2800~K~$>$~\tbol~$>$~650~K
\item Class III: \tbol~$>$~2800~K.
\end{itemize}
This system has the advantage of unifying Class 0 with the subsequent Classes; however, it still leaves out Flat Spectrum protostars that can fall within both the Class I and Class II ranges in \tbol. A more unified system became important as near-infrared sensitivity increased, because there were more detections of scattered light toward Class 0 protostars. The inclusion of emission from scattered light in the SEDs of Class 0 protostars (classified via L$_{\rm submm}$/L$_{\rm bol}$ or \tbol) can also make a system consistent with Class I. Case in point, L1527 IRS for which \tbol\ makes it a Class 0 protostar \citep{Ohashi2023}, but it is clearly detected shortward of 10~\micron\ (Figures \ref{L1527_multiwave} and \ref{L1527_SED}). Ultimately, the advent of the \textit{Spitzer Space Telescope} made the detection of Class 0 protostars between 3.6 to 8.0~\micron\ routine. We emphasize that for the vast majority of Class 0 protostars, their emission shortward of 10~\micron\ is dominated by scattered light \citep[e.g.,][]{Whitney2003TwodimensionalSequence} whether or not it appears point-like or is resolved into large bipolar nebulae. Direct detection of protostellar and/or inner disk emission is only possible if the orientation of the protostar is such that one is looking down the outflow cavity. 

The \tbol\ system is not without its own limitations. The definition for the Class 0 boundary in \tbol\ was ad hoc, simply because that was the point below which most known Class 0s from the L$_{\rm submm}$/L$_{\rm bol}$ criteria were found in \tbol.  A large comparison of the \tbol\ and L$_{\rm submm}$/L$_{\rm bol}$ criteria was shown in \citet{Dunham2014PPVI}, illustrating that there is reasonable agreement between the two definitions, though greater consistency could possibly be achieved by lowering the upper \tbol\ limit for Class 0. There is also a small, but not insignificant number of protostars that have L$_{\rm submm}$/L$_{\rm bol}$ consistent with Class 0, but \tbol\ consistent with Class I. Moreover, the accuracy at which both criteria can be calculated depends on the SED sampling with wavelength, which has improved dramatically in the past two decades, but is still a limitation. More broadly, the boundaries in \tbol\, and really the boundaries across both schemes, are observationally defined rather than being based on physical properties of the system. However, the intent of the L$_{\rm submm}$/L$_{\rm bol}$ criteria for Class 0s was intended to represent the point where M$_{\rm env}$ $>$ M$_*$.

As discussed previously, these classes are typically interpreted as an evolutionary sequence from heavily embedded protostar to a pre-main-sequence star. We present a graphical illustration of the protostellar classification scheme in Figure \ref{overview}, both with diagrams and example SEDs. A major weakness in treating these classifications as an evolutionary sequence, however, is that the classification of a source can be heavily influenced by viewing geometry. For example, an edge-on disk with no envelope could be obscured enough by the optical depth of the disk so that it has a rising SED, mimicking the properties of a Class I or even Class 0 protostar \citep[e.g.,][]{Crapsi2008}. Alternatively, a very young protostar viewed face-on could appear as a Flat Spectrum protostar. Indeed \citet{Sheehan2022a} conducted radiative transfer modeling of a sample of 97 protostars, fitting Atacama Large Millimeter/submillimeter Array (ALMA) images containing spatial information along with their SEDs, and found from the resulting best-fit models that the measured \tbol\ was correlated with the fitted inclination. That is to say that the classification of an object via spectral index or \tbol\ does not purely reflect changes in physical structure. Even the L$_{\rm submm}$/L$_{\rm bol}$ scheme is not immune to inclination because the measured \lbol\ can also be impacted by viewing geometry \citep{Whitney2003}. Furthermore, extinction toward the protostar from the surrounding molecular cloud, in addition to the self-extinction from the infalling envelope, can contribute to further steepening the SED slope, which can in turn make sources appear younger than they actually are \citep{Mcclure2010}.
In spite of these flaws, the system is  widely used and that fact will be reflected in our usage of the terms throughout this review. That said, readers should be mindful of these flaws when treating them as an evolutionary sequence, and we will revisit this topic in Section \ref{section:pathways}.

\subsection{Completeness of Protostar Samples}
The large area surveys at a few arcsecond resolution (or better) from the past two decades have resulted in a much more complete census of protostars \citep[e.g.,][]{Dunham2014}. The completeness of protostar censuses and robustness of the classifications depend on distance to the protostars and the amount of sky area covered with sufficient multi-wavelength data. \citep{Evans2009} estimated that their catalogs based on \textit{Spitzer} observations were $\sim$90\% complete to L$_{\rm bol}$ $\sim$0.08~\lsun\ for star forming regions within 300~pc. The \textit{Herschel} and submillimeter studies of the same regions only added a small number of protostars missed by \textit{Spitzer}. It is reasonable to conclude that 90\% of the protostars have been identified for star forming regions closer than 300~pc. 

The completeness of samples toward the more distant Orion (d$\sim$400~pc) \citep{Furlan2016} and Serpens-Aquila (d$\sim$436~pc) \citep{Pokhrel2023} regions are more difficult to quantify because the bright nebulae can limit the detectability of protostars, and Serpens-Aquila is near the Galactic Plane, making contamination and determining the bounds of the region an issue. \textit{Herschel} observations of Orion only added 15 new protostars to the original sample of 393 derived from \textit{Spitzer} data, but within regions of bright nebulosity and there is evidence for some hidden populations on the order of a few 10s of protostars \citep{Teixeira2016}. 
While those studies did not provide explicit completeness estimates, it is reasonable to assume that they are less complete, but not significantly less than 90\%. Thus, it is safe to conclude that there is not a large population of protostars within 500~pc that have gone undetected. 

Beyond 500~pc, the identification of individual protostar systems is significantly less robust given the modest resolution of \textit{Spitzer} and \textit{Herschel}. Detected sources in regions at $\sim$1~kpc and greater are often composed of multiple individual systems, making definitive classification challenging.

\subsection{Outline of this Review}
  We have organized this review largely by spatial scale, beginning at large-scales with cores within molecular clouds and progressing to disks. Along the way, we will also discuss the methods for characterizing envelope and disk structure, as well as what we know about protostar masses from their circumstellar material. At the end we summarize the main points of the review and finish by outlining what we believe are open questions that could be addressed in the forthcoming decade and beyond. Readers may notice that there are some areas that we do not cover in great detail, these include molecular clouds \citep{Heyer2015}, starless/pre-stellar cores \citep{Bergin2007}, magnetic fields \citep{Hull2019}, astrochemistry \citep{Jorgensen2020}, protoplanetary disks \citep{Andrews2020}. There are other recent reviews of these topics, but they will be covered insofar as they are relevant to particular topics that we cover regarding protostars.

\section{From Clouds to Cores to Disks}
The vast majority of protostars are found within molecular clouds \citep{Dunham2014},  though there are some protostars that are forming in truly isolated dense cores \citep[e.g., B335;][]{Stutz2008}. It has been known since the time of early optical studies of dark clouds that molecular clouds are threaded with filamentary structure \citep{Barnard1919}, and observations from the \textit{Herschel Space Observatory} solidified our understanding of protostellar cores forming inside filaments within molecular clouds \citep{Andre2014}. Kinematic studies of filaments have even found flows along filaments feeding dense cores \citep{Hacar2011}. Although we will concentrate primarily on the protostellar cores, it is important to remember that most cores do not exist in isolation and that they retain a connection to their molecular clouds that can make well-defined boundaries difficult to establish.

\subsection{Observations of Protostellar Core and Envelope Structure} 
Assumptions of symmetry have dominated the study of protostellar cores for decades. This was driven, in part, by the fact that the resolution of instruments capable of observing cores and envelopes only offered a resolution of 1000s of au at best, and consequently there were few hints of cores and envelopes deviating from symmetry. However, over the the past two decades there have been substantial advances in our knowledge of envelope structure that have begun to shed these simplistic assumptions.

\label{section:core_struct}
The pioneering work on protostellar cores using high-density molecular line tracers \citep{BensonMyers1989,Myers1991,Caselli2002} found early evidence that some cores exhibited significant asymmetry, as the location of the mid- to far-infrared sources detected were sometimes located off-center or offset from the molecular line emission.  However, the structure of the molecular line emission was different depending on the tracer observed. Thus, envelope structure itself could not be reliably probed without an independent tracer of the dense gas not influenced by chemistry.

The advent of sensitive submillimeter continuum imaging using bolometer arrays led to the characterization of core and envelope structures using dust emission. Large samples of envelopes were observed at higher resolution than previously possible and they were found to exhibit some mild asymmetries \citep{Shirley2000}. A limitation of dust emission as a probe of protostellar core and envelope structure, however, is that the emission of dust is biased by temperature, so warmer regions of envelopes appear brightest in sub-millimeter maps. This may give a false impression of symmetry because the temperature profiles are approximately radial. This is evident in Figure \ref{L1527_multiwave} where the brightest emission in the 450~\micron\, 850~\micron, and 3~mm images appears symmetric, but because of the high sensitivity of the latest 450 and 850~\micron\ maps, extended emission at relatively low surface brightness compared to the peak emission is evident.

An alternative technique, mid-infrared extinction mapping, has been used to greatly advance the study of distant clouds. The dense, dusty clouds in the Galactic plane extincted background emission efficiently at 8~\micron\ revealing an exquisite filamentary structure \citep{Perault1996,Carey1998,Ragan2009}. Dust extinction mapping provides an alternative method for characterizing cloud and core structure that complements mapping of thermal dust emission.

\begin{figure}
\minipage{0.33\textwidth}
\includegraphics[width=\linewidth]{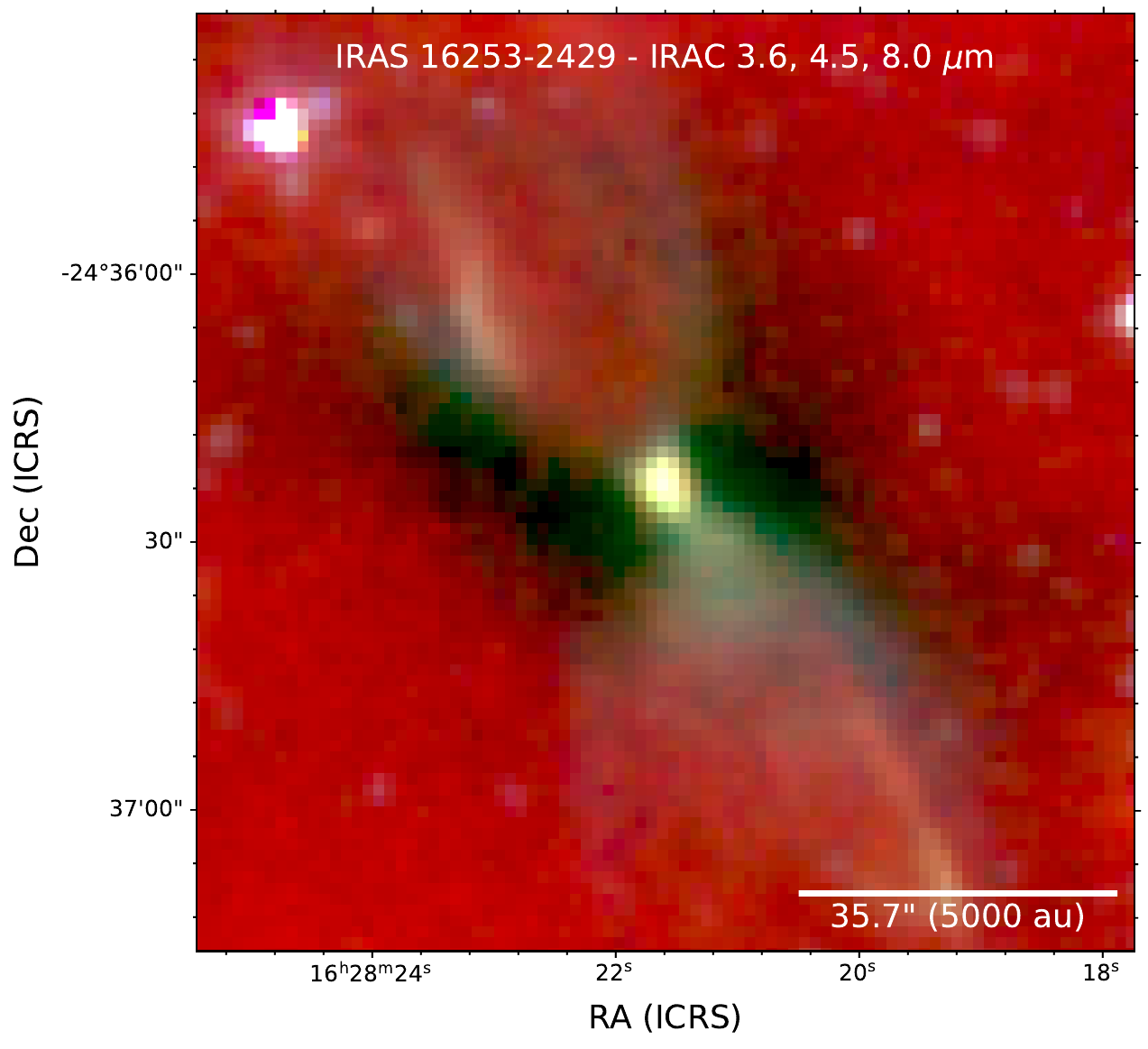}
\endminipage\hfill
\minipage{0.33\textwidth}
\includegraphics[width=\linewidth]{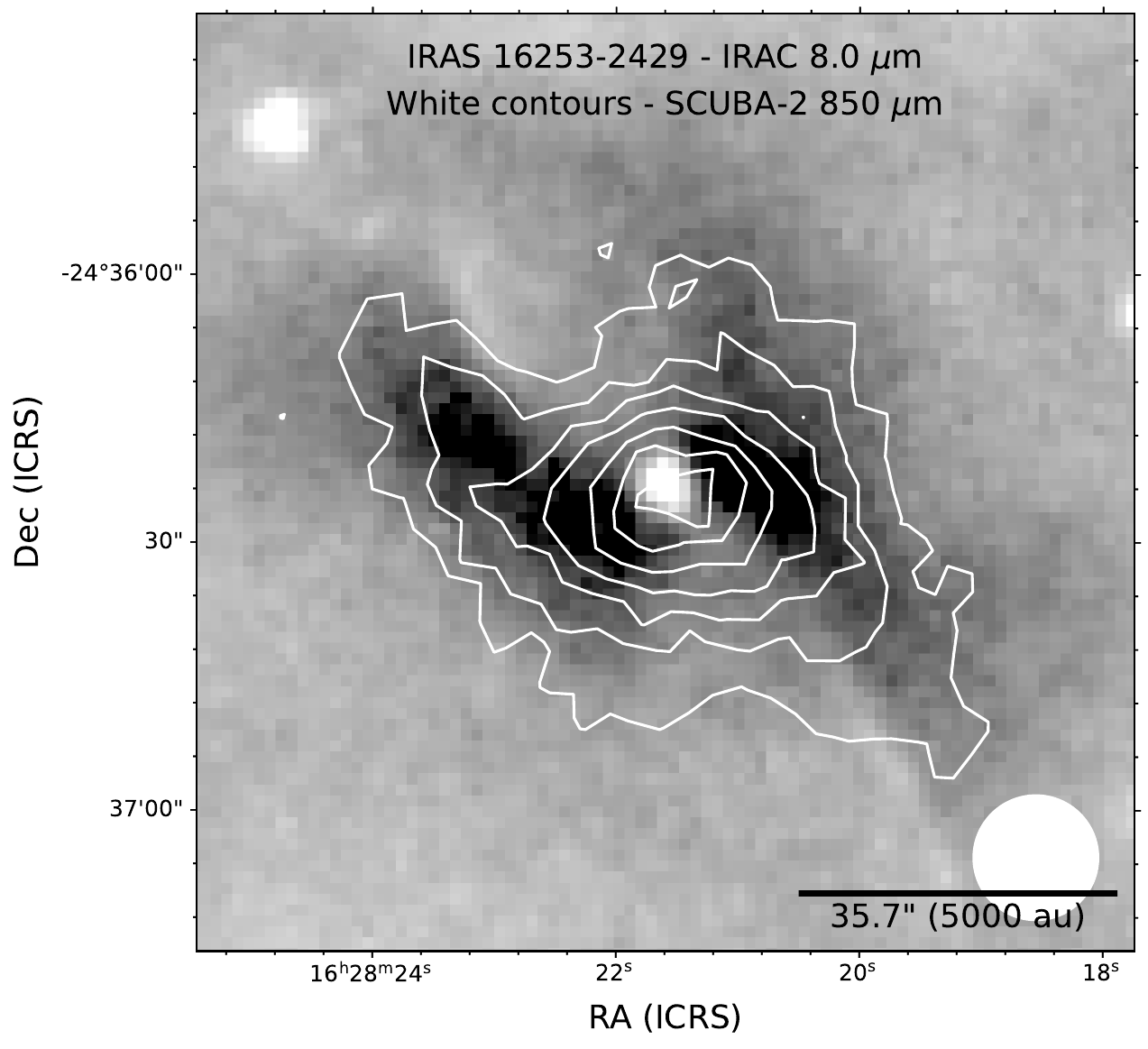}
\endminipage\hfill
\minipage{0.33\textwidth}
\includegraphics[width=\linewidth]{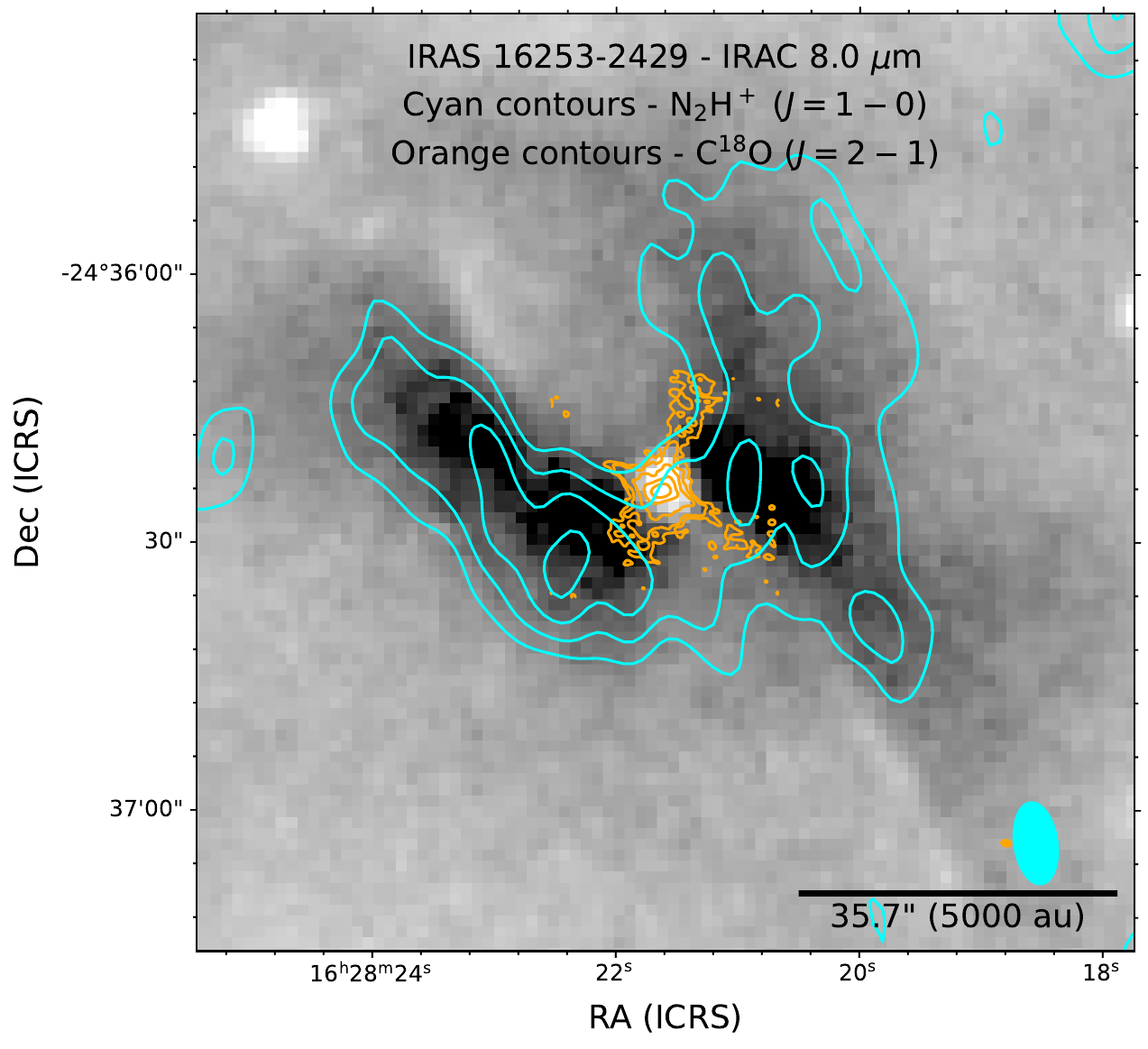}
\endminipage\hfill

\minipage{0.33\textwidth}
\includegraphics[width=\linewidth]{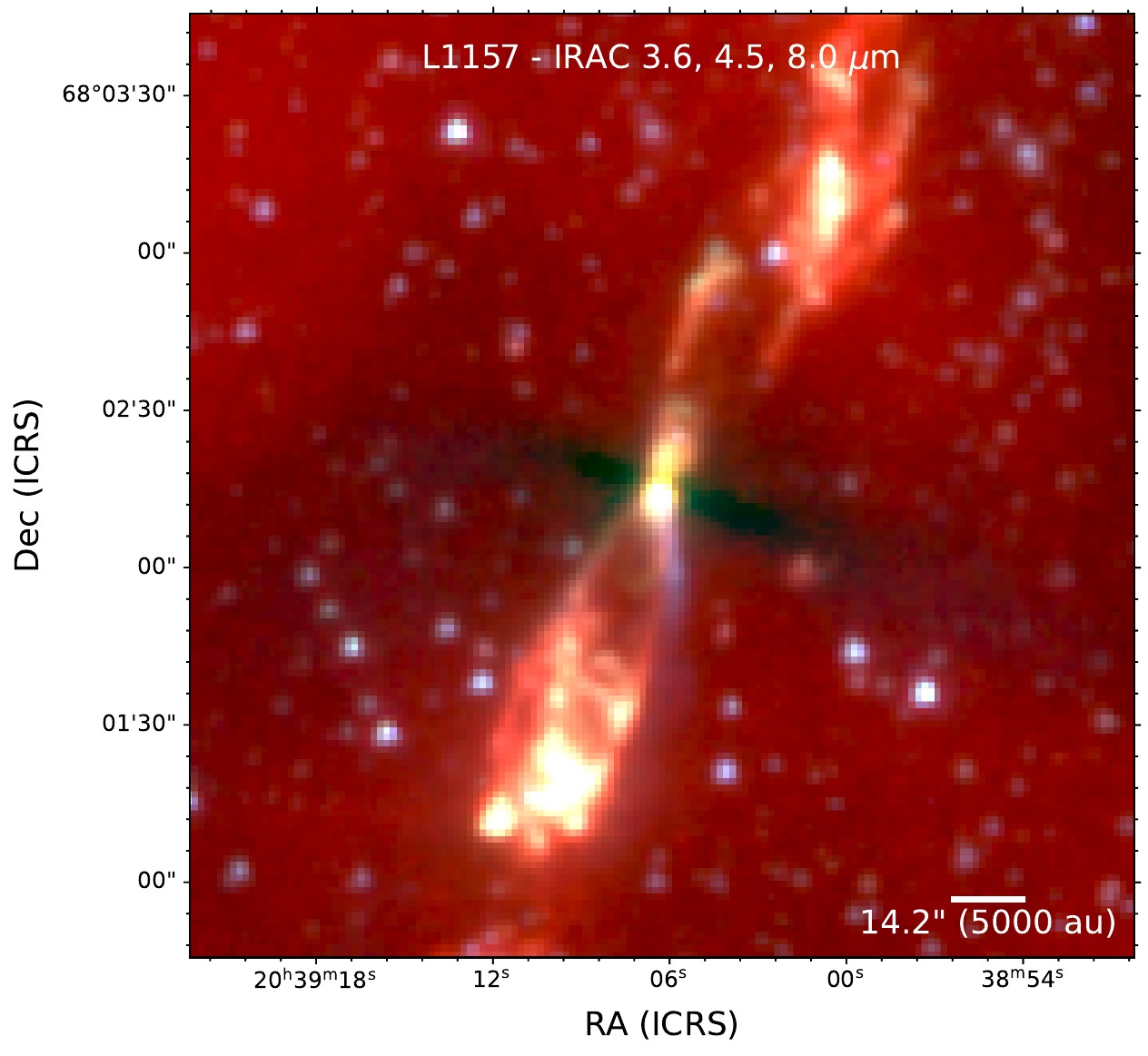}
\endminipage\hfill
\minipage{0.33\textwidth}
\includegraphics[width=\linewidth]{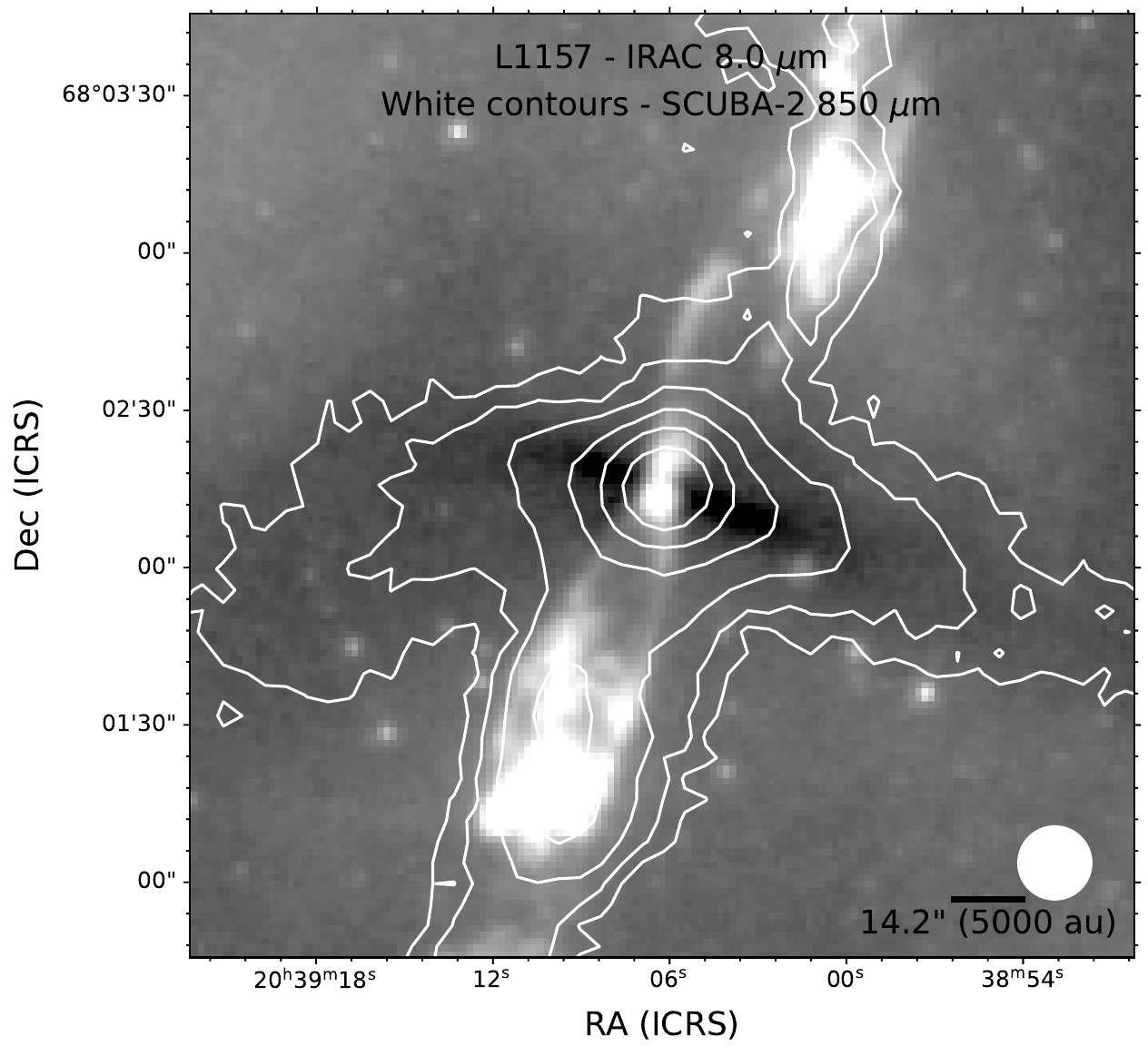}
\endminipage\hfill
\minipage{0.33\textwidth}
\includegraphics[width=\linewidth]{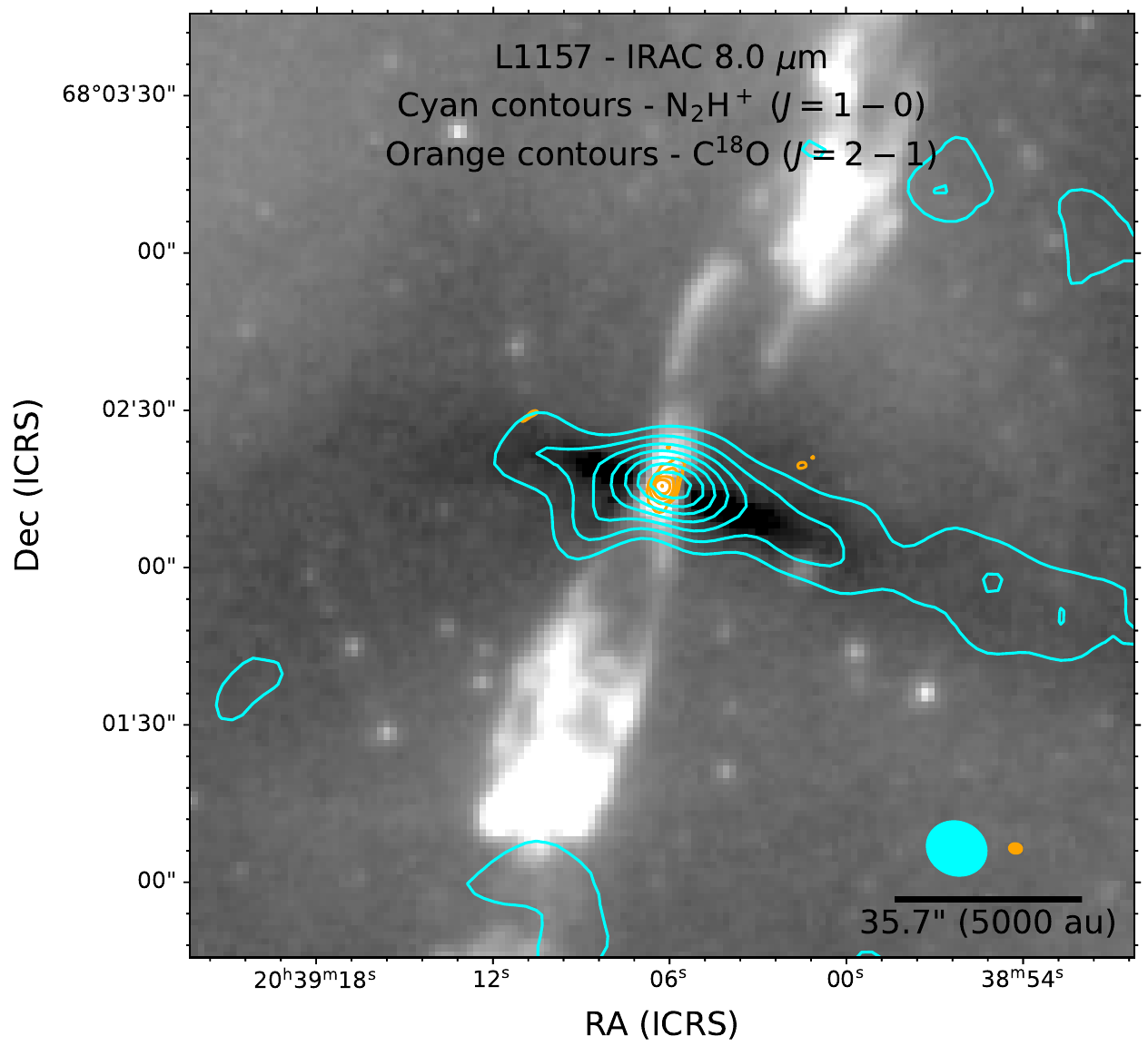}
\endminipage\hfill

\caption{Multi-wavelength and multi-tracer images toward Class 0 protostars IRAS 16253-1229 (top row) and L1157 (bottom row). The left panels show a 3-color image from IRAC 3.6, 4.5 and 8.0~\micron\ imaging from \textit{Spitzer}, the middle panels show the 8.0~\micron\ image with SCUBA-2 850~\micron\ contours overlaid, and the right panels show the 8.0~\micron\ image again with \nthp\ ($J=1\rightarrow0$) integrated intensity contours and \cateo\ ($J=2\rightarrow1$) contours overlaid. The relatively symmetric envelope for IRAS 16253-2429 and the highly flattened envelope for L1157 are evident in the 8~\micron\ absorption images and the absorption is also well-traced by the SCUBA-2 850~\micron\ image and \nthp. The \cateo\ only traces the central regions of the system where temperatures are high enough for CO ice to be sublimated from the dust grains. Note that there is some contamination from \twco\ to the 850~\micron\ continuum, which is why it also traces the outflow for L1157. The beams in the middle and right panels are shown in the lower right with the same color as their contours. }
 \label{IRAS16253_L1157}
\end{figure}

This technique was also found to be quite fruitful for nearby protostellar cores/envelopes when the Galactic background emission was sufficiently bright and confusion with neighboring YSOs and nebulosity was minimal. It was thereby revealed that envelopes could exhibit a wide variety of structures, from quasi-symmetric, to highly flattened, to highly asymmetric \citep{Looney2007,Stutz2008,Tobin2010}. Thus, it was convincingly shown that the cores and envelopes tend to be asymmetric.
We show two examples in Figure \ref{IRAS16253_L1157}; IRAS 16253-2429 appears the most symmetric of the sample published in \citet{Tobin2010}, while L1157 appears highly flattened or filamentary, with a wide tail curving to the southeast. Examples of recent 850~\micron\ mapping also appear in Figure \ref{IRAS16253_L1157}, and their observed structure in thermal dust emission often agrees very well with the structures detected using 8~\micron\ extinction mapping from \textit{Spitzer}.

Space-based submillimeter mapping from the \textit{Herschel Space Observatory} provided a much less biased view of dust emission, enabling the colder, low-surface brightness dust emission to be detected alongside the brighter, warmer regions alike \citep[e.g.,][]{Stutz2010}. 
While resolution was still limited, the capacity for large samples at high sensitivity in nearby regions led to many envelopes and cores being mapped, with many having substantial deviations from symmetry \citep{Launhardt2013,Sadavoy2018}. Some examples of \textit{Herschel} mapping at 70, 100, and 160~\micron\ are shown in Figure \ref{L1527_multiwave}.

Interferometers have also become much more sensitive over the past two decades, enabling the continuum emission of the inner envelopes to be examined at significantly higher resolution. That said, sensitivity and interferometric filtering affect the resultant images, and they too are affected by the radially decreasing temperature profile of protostars. Thus, early continuum images appear approximately symmetric and centrally peaked \citep{Looney2000,Looney2003}. However, in the case of L1157, the envelope continuum emission does trace the flattened structure \citep{Stephens2013} at low surface brightness. While it can be difficult to map the extended envelope continuum emission, molecular lines like \nthp\ can be detected more readily. Figure \ref{IRAS16253_L1157} shows that the asymmetric dust extinction traced at 8~\micron\ is well-traced by \nthp\ \citep[and \nht;][]{Tobin2011}. \nthp\ cannot trace the envelope all the way to the disk, however, because \nthp\ is destroyed by CO, which sublimates from the dust grains at T~$\gtrsim$~20~K. Then for both IRAS 16253-2429 and L1157, the \cateo\ emission is shown to peak at the protostar position due to the higher dust temperatures at smaller radii.

Both early and current interferometers have provided constraints on the radial density profiles of infalling envelopes. 
Multiple studies found that the envelopes are often more consistent with a profile of $\rho$~$\propto$~r$^{-2}$, than $\rho$~$\propto$~r$^{-1.5}$ \citep{Looney2003,Jorgensen2005,Jorgensen2009,Maury2019}. These results are in contrast to an inside-out collapse scenario \citep{Shu1977}, where the density profile of the infalling material is expected to be the latter, whereas the measured values favor the steeper density profile predicted by \citet{Larson1969}. A limitation of most analyses of envelope density profiles, however, is that they utilize circularly or elliptically-averaged visibility amplitude profiles. Applying circular averaging to a fundamentally asymmetric or elongated structure can give the signature of a steeper density profile even if none is present due to averaging over empty area, so furthering this analysis with two-dimensional fits may be needed in the future.

An additional confounding factor with regard to observations of envelope structure can be the outflows from the protostars. Outflows may wander and precess, and perhaps greatly change their position angles during the course of protostellar evolution \citep[e.g.,][]{Frank2014}. As such, they affect the structure of observed protostellar cores by removing portions of their envelope material. Additionally, the excavated outflow cavities will then be illuminated by the protostar and accretion disk, thereby increasing the dust temperatures along the outflow and consequently further increasing the intensity of the dust emission. In many of the cases shown in \citet{Tobin2010}, the asymmetric dust emission does not appear to have been plausibly impacted by outflows given that the dense material was typically extended orthogonal to the known outflow directions. Such impacts in the past, however, are difficult to rule-out completely, but, there is little evidence for most outflows to have dramatically changed their directions in the past several thousand years.

\subsection{Core Kinematics}

The molecular emission from protostellar cores not only traces core structure, but also the gas kinematics. The dense gas tracers of choice for probing protostellar cores are still \nthp, \nht, and their deuterated counterparts, as the critical densities of \nthp\ and \nht\ are $\sim$10$^5$~cm$^{-3}$ and $\sim$10$^3$~cm$^{-3}$, respectively. Thus, \nht\ can be detected in emission at lower densities and farther from the core centers than \nthp, but its formation requires higher densities. The kinematic structure obtained from \nht\ and \nthp\ are very similar for both the centroid velocities and line widths \citep{Johnstone2010,Tobin2011}, but there may be minor systematic deviations that only become apparent with high sensitivity and resolution \citep{Pineda2021}. 

Modern, large single-dish telescopes are able to trace core structure and kinematics out to larger radii and lower column densities, while having a smaller beam than earlier studies \citep[e.g.,][]{Pineda2010,Tobin2011}. As such, modern interferometers are able to trace the internal envelope kinematics at $\lesssim$5\arcsec\ resolution. Such spatially-resolved observations of cores and envelopes are found to closely trace the envelope structures probed by \textit{Spitzer} at $\sim$2\arcsec\ resolution in 8~\micron\ extinction \citep{Chen2007,Tobin2011,Hsieh2019}. The column densities of flattened envelopes indicate that they are filament like rather than sheet-like along the line of sight \citep{Tobin2010}.

\begin{figure}
\includegraphics[width=\linewidth]{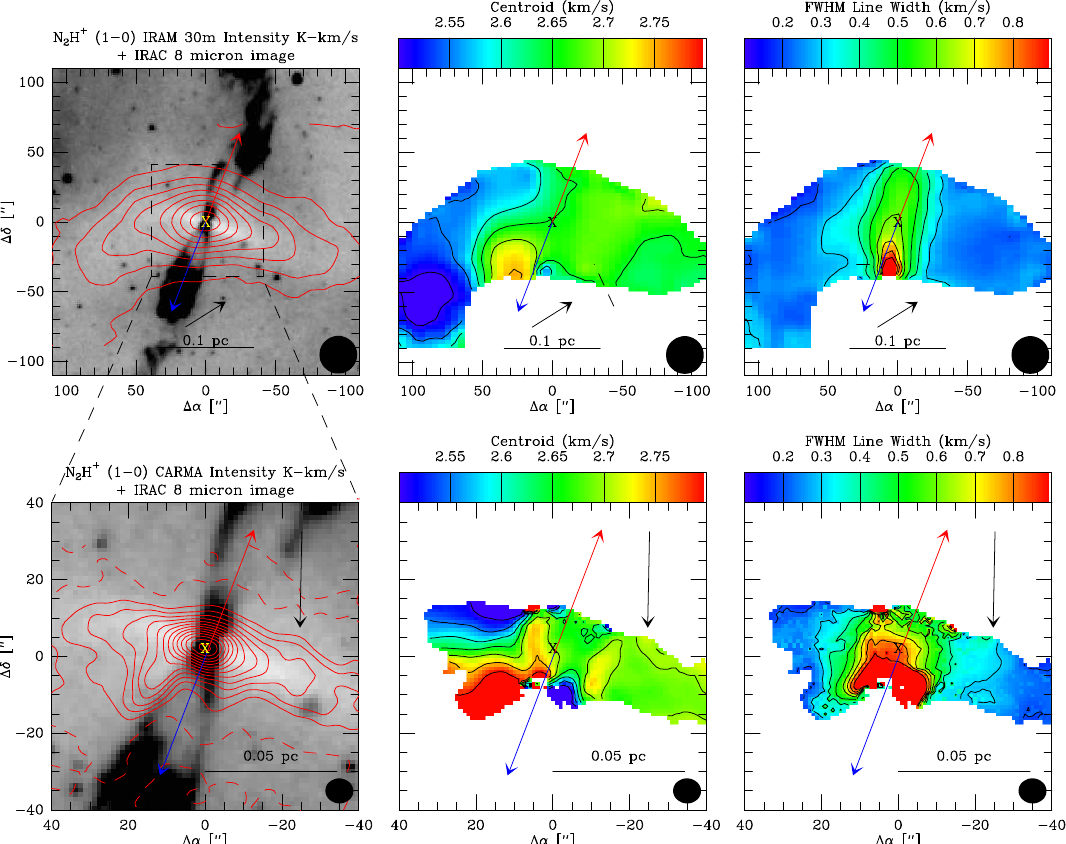}
\caption{Single-dish (IRAM 30m) and interferometer (CARMA) observations of \nthp\ toward the Class 0 protostar L1157. The single-dish data are shown in the top row, while the interferometer data are shown in the bottom row. The left column shows the \nthp\ integrated intensity contours overlaid on the IRAC 8~\micron\ image, the middle column shows the line center velocity of the \nthp\ emission (note the narrow range of velocities), and the right panel shows the FWHM linewidth of \nthp. The blue and red arrows denote the outflows and their orientation along the line of sight, the 'X' marks the position of the protostar, and the observing beams are shown in the lower right corners of each panel. Reproduced from \citet{Tobin2011}.}
\label{L1157_kinematics}
\end{figure}

The kinematic structure of the envelopes is examined through velocity centroid maps along the line of sight, derived from molecular line observations. Previous single-dish measurements of velocity gradients were interpreted as tracing solid-body rotation in the protostellar cores \citep{Goodman1993} due to the frequent appearance of linear velocity gradients. A single-dish observation is shown in Figure \ref{L1157_kinematics} where, at $\sim$30\arcsec\ resolution, the velocity gradient appears relatively smooth across the core. However, higher resolution views of the gas kinematics paint a more complex picture wherein the velocity structure is more likely tracing a combination of projected infalling motions, residual turbulence, and sometimes outflow sculpting, rather than pure rotation \citep{Dib2010,Tobin2011,Tobin2012a,Galametz2020}. The increased kinematic detail shown in the interferometer observations in Figure \ref{L1157_kinematics} demonstrates significant departures from well-ordered kinematic structure. The lack of well-organized rotation being a dominant feature of core kinematics when viewed at higher resolution has resolved some tensions. When velocity gradients are interpreted as rotation, they often suggested the formation of excessively large rotationally-supported regions ($\sim$1000~au) from conservation of angular momentum (in the absence of processes like magnetic braking \citep{Tobin2012a}. Instead, \citet{Pineda2019} performed a detailed analysis on the line of sight velocity maps from three sources to derive their specific angular momenta, inferring upper disk size limits of $\sim$60~au, even though the kinematics did not clearly resemble rotation in all cases.
\begin{marginnote}[]
\entry{IRAM}{Institut de radioastronomie millimétrique}
\entry{CARMA}{Combined Array for Research in Millimeter-wave Astronomy}
\end{marginnote}

Cores and envelopes have consistently been found to have line widths larger than the expected thermal line widths of \nthp\ or \nht\ \citep{Goodman1993,Caselli2002,Chen2007}. This has been interpreted as subsonic turbulence within the cores, and it has been argued that the transition from supersonic turbulence to subsonic turbulence could be a potential diagnostic for the boundary of cores \citep{Pineda2010}. Transitions to coherence have been observed for a larger number of cores, thanks to large-scale single-dish surveys \citep{Friesen2017, Chen2019a}.

\subsection{Core Fragmentation}

A telltale sign of core fragmentation is the oft observed multiple protostar systems having separations from 100s to 1000s of AU \citep{Looney2000,Chen2013,Tobin2022}.
A few studies using molecular lines have even been able to discern the relative line of sight velocities for the individual protostars, finding that some are bound while others are close to being unbound \citep{Lee2015}. 

Rotation of cores has often been looked to as a mechanism for fragmentation in the past, with measurements of core velocity gradients being relatively straightforward to convert into $\beta$, the ratio of rotation to gravitational energy. Models have shown that strong core rotation could indeed lead to fragmentation into a multiple system, and studies have found that $\beta$ $\sim$0.01 was frequently a dividing line between protostars with observed companions and those without \citep[e.g.,][]{Chen2012}. That said, while some cores may indeed be rotating (or have some rotational component), the detailed kinematic maps discussed in the previous section complicate the interpretation of solid-body rotation. Moreover, the fact that outflows are randomly oriented, even for protostars within the same core \citep{Lee2015,Stephens2017}, suggests that fragmentation is not simply due to rotation, which would result in correlated angular momentum axes (and hence outflows axes).

The most appealing alternative to rotation is turbulent fragmentation \citep{Padoan2002,Offner2010}. In this mode of fragmentation, the turbulent velocities of the molecular cloud, which lead to core formation itself, can cause a further cascade of fragmentation within cores. The lack of well-ordered velocity patterns along with the subsonic turbulence within cores may provide the seeds for multiple protostars to form within cores. 
An additional route that has been proposed is that the asymmetric core structures that result from their formation within a turbulent medium can also form multiple protostars via gravitational focusing, in which case different portions of the core can collapse independently from each other \citep{Tobin2010}. 
These asymmetric core structures within a turbulent medium may further manifest at smaller scales during collapse, within the envelopes feeding the nascent disk around the protostars, and are often referred to as streamers.

\subsection{The Nature of Streamers}

\begin{figure}
\minipage{0.49\textwidth}
\includegraphics[width=\linewidth]{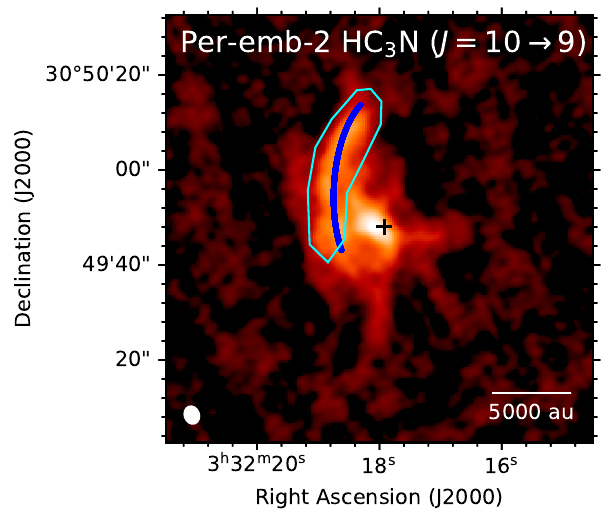
}
\endminipage\hfill
\minipage{0.49\textwidth}
\includegraphics[width=\linewidth]{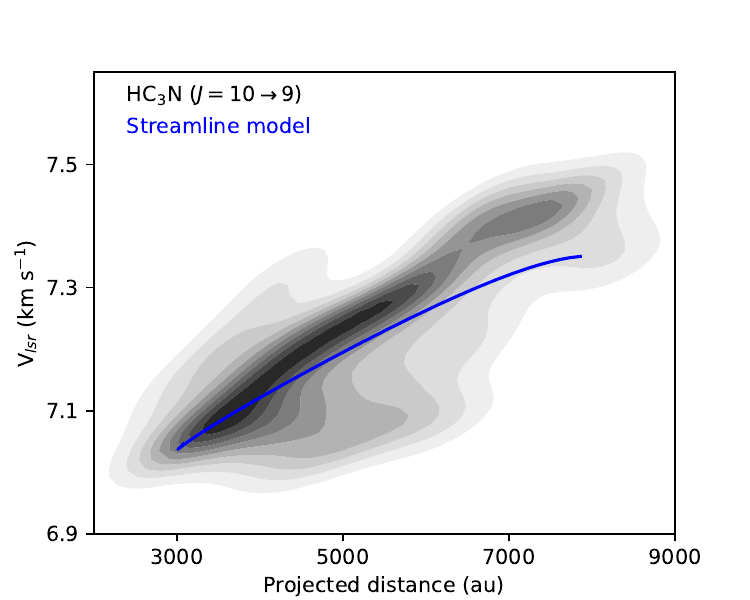}
\endminipage\hfill
\minipage{0.48\textwidth}
\includegraphics[width=\linewidth]{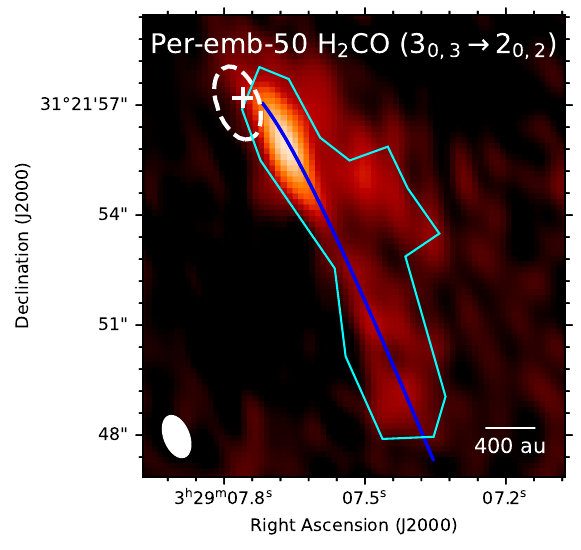}
\endminipage\hfill
\minipage{0.49\textwidth}
\includegraphics[width=\linewidth]{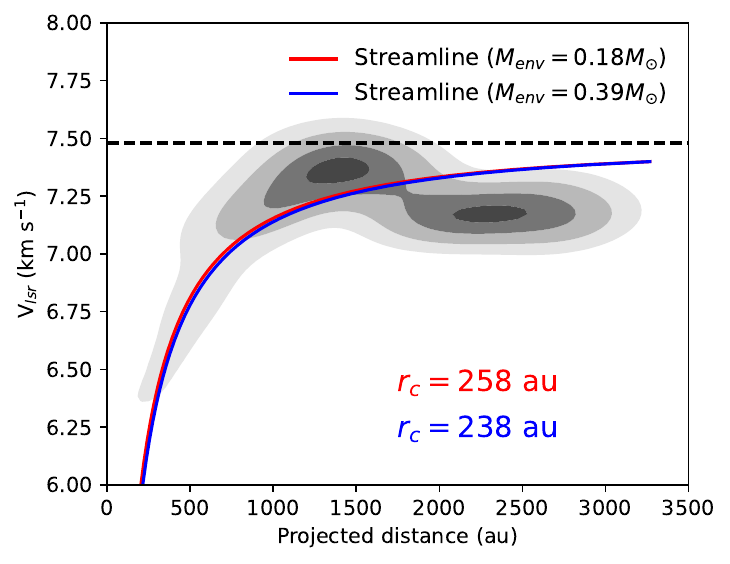}
\endminipage\hfill
\caption{
Examples of large (top) and small-scale streamers (bottom) toward the Class 0 protostar Per-emb-2 from \citet{Pineda2020} and the Class I protostar Per-emb-50 from \citet{Valdivia2022}, respectively. Per-emb-2 (top) shows large-scale emission in HC$_3$N whose kinematics are consistent with infall from the outer envelope to the inner envelope. This streamer is beyond the the scale of the envelope as viewed in \nthp.  While Per-emb-50 (bottom) shows emission from within a few thousand au that is shown to be infalling toward the disk (white dashed ellipse).
The model streamlines plotted in the right panels for both sources over the position-velocity diagram from the data are calculated using the formulation of rotating collapse for a finite object from \citet{Mendoza2009}. The fits show that the material is consistent with infalling gas for both the 10000~au and 3000~au streamers.
}
\label{streamers}
\end{figure}

Streamers have recently become a key topic in core structure. The term implies a structure that is relatively confined in its azimuthal width and its presumed depth into the plane of the sky. The same term has been used to describe structures that are occurring on $\sim$10000~au to $\sim$1000~au scales (see Figure \ref{streamers} for examples), despite the possibly disparate processes and timescales that may be involved.

 \citet{Pineda2020} reported the first instance of a $\sim$10000~au streamer, traced by HC$_3$N. Such large-scale streamers may be related to the asymmetric envelopes that have been seen in absorption at 8~\micron\ toward some protostars (see Section \ref{section:core_struct}); however, the asymmetric envelopes tended to be well-traced by \nht\ and \nthp\ \citep{Tobin2011}, and indeed \nthp\ toward this protostellar core is only detected at several thousand au radii. Thus, the asymmetric structure traced by these large-scale streamers is likely probing even lower-density material, and \citet{Pineda2020} suggested that large-scale streamers connect envelopes to the molecular cloud. As such, large-scale streamers play an important role in tracing the flow of material from the molecular cloud into the envelope(s) and disk(s), and gaining a more complete picture of their properties will require a multi-scale and multi-tracer approach to continuously characterize the flow of gas as one tracer loses effectiveness and another becomes more effective. 

Smaller-scale streamers are becoming frequently observed toward protostars on $\sim$1000~au scales \citep[][e.g. see Figure \ref{streamers}]{Valdivia2022,Murillo2022,Valdivia2023,Thieme2022,Mercimek2023,Flores2023}. The detected small-scale streamers are inconsistent in their detectability; some are only apparent in certain molecular tracers, while others are detected with multiple species. The inconsistent detection makes it unclear whether streamers are actual density enhancements in an otherwise symmetric inner envelope, as opposed to transient chemical signatures or elevated temperatures. Streamers are expected to be embedded within a more symmetric envelope (e.g., Figure \ref{overview}). Otherwise, if the envelope was only composed of high density streamers that did not fully surround the protostars, there would be more protostars detected at optical and/or X-rays wavelengths, which would likely be classified as Flat Spectrum protostars, depending on foreground extinction. Surveys find very few Flat Spectrum or optically-detected protostars that are envelope dominated \citep{Furlan2016, Federman2023}.

Some streamers appear to follow outflow cavity walls in projection \citep{Murillo2022,Thieme2022,Mercimek2023,Flores2023}. This association presents the possibility that some streamers are really temperature-enhanced gas within a possibly more symmetric envelope, rather than coherent overdensities. This does not mean that the features are part of the outflow. Indeed, kinematic analyses indicate that streamers on the whole have kinematics that imply infall and not outflow. But the appearance of this material as a streamer could be due to the temperature enhancement of the gas located near the outflow cavities that is directly illuminated. We do emphasize, on the other hand, that there are also nearly as many streamers that are most likely outside the influence of the outflow cavity and may represent density enhancements \citep[][e.g., see Figure \ref{streamers}]{Valdivia2022,Kido2023,Murillo2022}.

The further characterization of streamers is necessary to understand their importance to the star and planet formation process. The extent to which streamers feed disks unevenly is important because they could trigger instabilities in the disk \citep{Lesur2015,Kuznetsova2022}, possibly leading to the formation of disk substructure, the formation of companions via gravitational instabilities, and/or give rise to outbursts of high mass accretion from the disk to the protostar. Still, at this time our understanding of streamers is in its infancy, and the uncertainties in their nature will need to be rectified before their impact on the star and planet formation process can be fully understood.

\subsection{Kinematics of Disk Formation}

While it is clear that the envelope kinematics on scales $>$2000~au are not likely to trace pure rotation, interferometers provide the ability to directly characterize the kinematics from $\sim$1000~au down to the disk. The warmer temperatures at radii $\lesssim$2000~au release CO into the gas phase, making \thco\, \cateo, and \csevo\ the tracers of choice. Early interferometers had difficulty tracing envelope and disk kinematics with spatial resolution $<$100~au, outside of some of the nearest protostars \citep{Jorgensen2009,Tobin2012,Harsono2014}. The advent of ALMA, and to some degree NOEMA \citep[e.g.,][]{Maret2020}, has enabled the inner envelope kinematics to be probed efficiently for a much larger sample of protostars \citep[e.g.,][]{Murillo2013,Yen2014,Aso2015,Yen2017,Ohashi2023}, and at resolutions that can often probe the transition from the disk to the envelope.

A full picture of the kinematics from $>$1000~au down to the disk has not yet been unraveled, as it is complicated by the need for multi-scale observations of multiple molecular lines spanning the various chemical transitions in a core-envelope-disk system \citep{Tychoniec2021WhichSources}. 
One highly sought after kinematic signature is the flattening of the specific angular momentum with decreasing radius, which would indicate that angular momentum of the infalling material is conserved. This signature would be most prominent in the context of the inside-out collapse model with initially solid-body rotation of the collapsing core; however, the region of conserved angular momentum could be present with a variety of collapse and initial rotation profiles. Recent results, however, have not yet concretely identified this signature. If the observed envelope kinematics are interpreted as rotation, \citet{Pineda2019} see a decrease in specific angular momentum that is very similar to solid-body rotation, but if the kinematics are mix of infall and rotation this signature could be misleading. In a separate work, \citet{Yen2017} find that the conserved angular momentum region is in the inner envelope for most protostars, and many protostars appear to agree with the expectation for inside-out collapse of a core with initially solid-body rotation. However, the kinematics are only traced to radii larger than $\sim$1000~au for a small number of their sources.

The transition from the infalling/rotating envelope to a rotationally-supported disk has been a matter of intense study in the last decade. Theoretical studies of disk formation within an infalling envelope provide a framework for understanding the flow of material from the envelope to the disk \citep{Ulrich1976, Terebey1984}. These models do not specifically describe how material is incorporated into the disk, other than that it falls-in until it reaches the centrifugal radius, where it can go into orbit around the protostar. The material entering the disk is expected to result in a shock, which may cancel out (at least partially) the velocities in the vertical and radial direction. 

\begin{figure}
\minipage{0.49\textwidth}
\includegraphics[width=\linewidth]{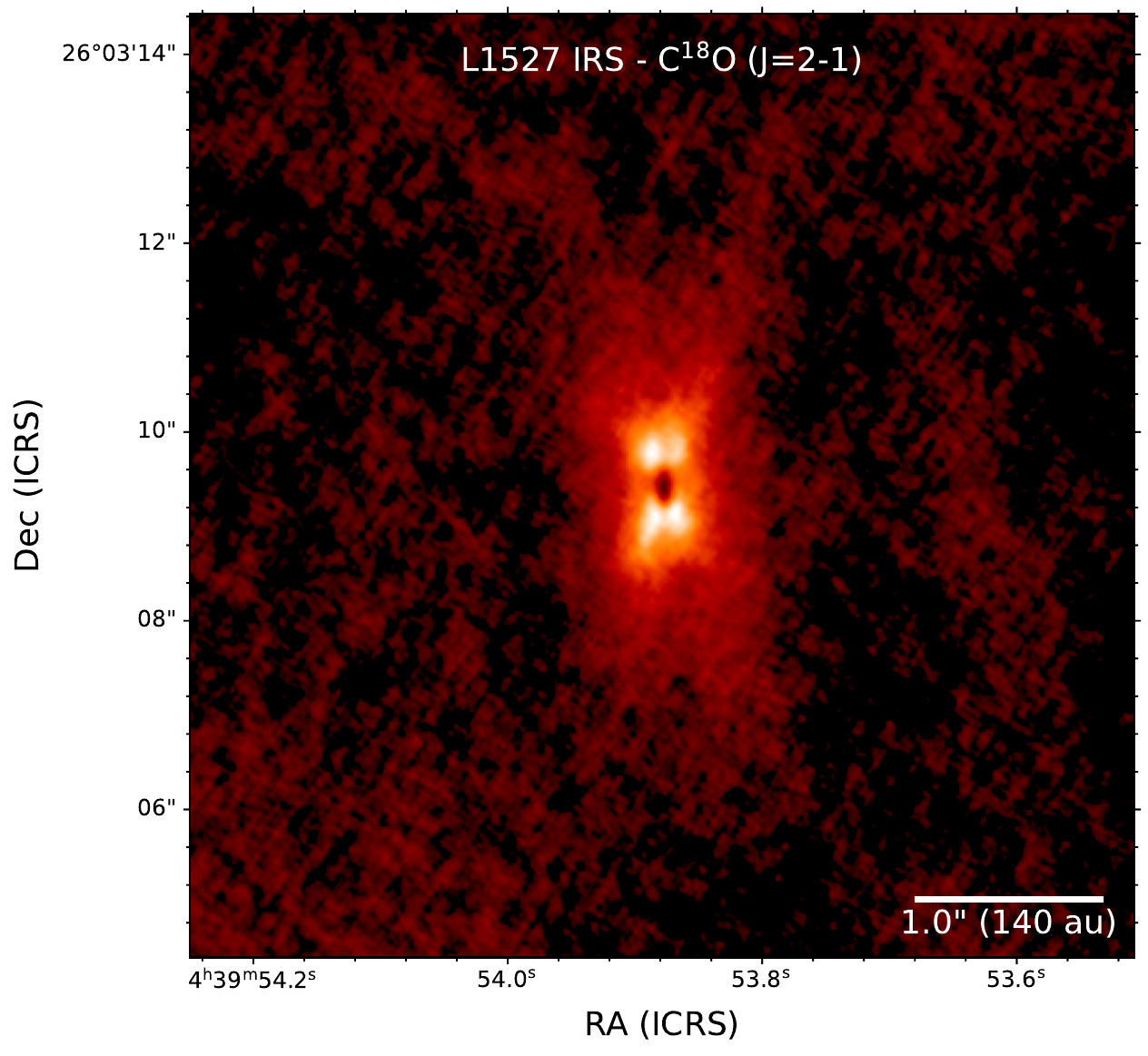}
\endminipage\hfill
\minipage{0.49\textwidth}
\includegraphics[width=\linewidth]{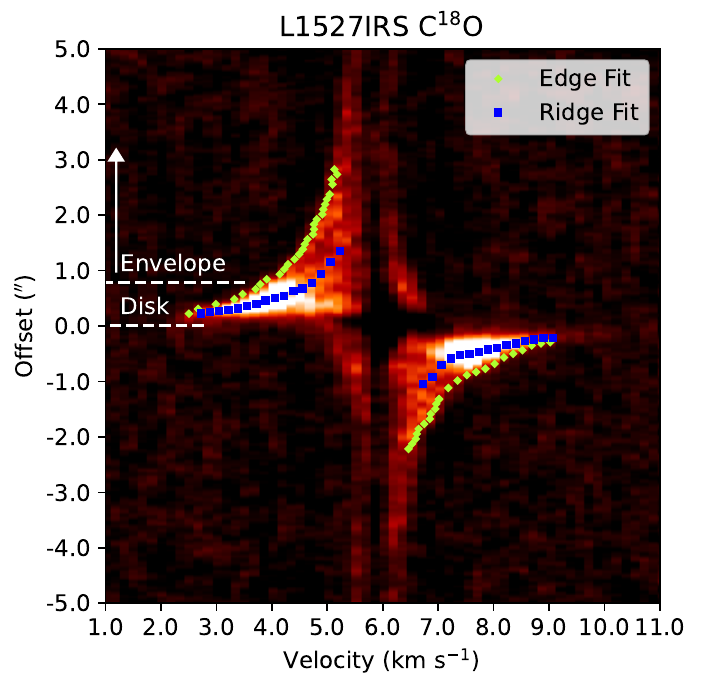}
\endminipage\hfill
\caption{ALMA \cateo\ ($J=2\rightarrow1$) molecular line observations toward L1527 IRS. The integrated intensity emission is shown in the left panel and the kinematics are shown on the right using a position-velocity diagram, showing the inner envelope and disk. The position-velocity diagram in the right panel clearly shows the kinematics associated with the disk forming region of the envelope where velocities increase toward smaller radii, eventually reaching Keplerian, rotationally-supported velocities. The central depression is the result of optically thick dust absorbing the \cateo\ emission. The green and blue points (edge and ridge fits, respectively; see Section 4) in the right panel represent fits of the rotation curves of the \cateo\ emission and indicated a disk of $\sim$100~au radius and a protostar mass of $\sim$0.4~\msun. The data were originally published in \citet{vanthoff2023}; see Section 4 for a description of the Edge and Ridge fitting methods.
}
\label{L1527_kinematics}
\end{figure}

The Class 0 protostar L1527 IRS shows a proto-typical edge-on disk in both the gas (Figure \ref{L1527_kinematics}) and dust emission (Figure \ref{L1527_multiwave}). The protostar shows a clear increase of the linewidth in the envelope outside the disk ($>$1\arcsec) due to the increased envelope rotation, and at $<$ 1\arcsec\ the linewidth continues to increase toward smaller radii in the position-velocity (PV) diagram due to the rotation of the disk. Thus, L1527 IRS shows a `classical' view of a forming protostellar disk within an infalling/rotating envelope. \citet{vanthoff2023} fitted its rotation curve, finding a $\sim$100~au radius rotationally-supported disk, and outside the disk the rotation curve is best described by an infalling/rotating envelope that is conserving angular momentum. 

The ideal view of the inner envelope and disk of L1527 IRS has attracted numerous studies of molecular emission from the disk forming region. Some studies have found evidence of chemical transitions from the envelope to disk. For instance, CCH and c-C$_3$H$_2$ become undetectable at the radius of the disk in L1527 IRS, while SO becomes bright. L1527 IRS thus far provides the cleanest picture of this chemical change from the envelope to the disk. To explain this chemical transition, \citet{Sakai2014a} utilized a ballistic model for infall, where infall continues and angular momentum is conserved inside the centrifugal radius. Then at half the centrifugal radius, all the infalling motion is fully translated to super-Keplerian rotation. The authors describe this as the `centrifugal-barrier' model, and suggest that the `centrifugal-barrier' is the cause of the chemical change by shock-heating the gas.

The `centrifugal-barrier' model, however, presents some difficulties; the infalling gas within the centrifugal radius is not a collisionless system, and indeed a centrifugal barrier is not produced in hydrodynamic or magneto-hydrodynamic models of protostellar collapse and disk formation \citep{Yorke1999,Zhao2016,Jones2022,Shariff2022}. Moreover, the super-Keplerian rotation yields protostar masses that are systematically lower relative to other work \citep[e.g.,][]{Yen2017}. 
Recent analyses of the centrifugal barrier point out that in order for the ballistic approximation to be valid, the pressure gradient must be negligible compared to radial advective acceleration \citep{Shariff2022}, and this criterion is not met in typical conditions. Thus, the `centrifugal-barrier' seems to be an artificial result of the simple physics employed. As such, any shock that is occurring is most likely to happen around the centrifugal radius, where the radial velocity component must be significantly reduced as material arrives at the disk. Therefore, while the transition from envelope to disk will not be completely smooth, a centrifugal barrier does not seem realistic. Furthermore, new observations of the L1527 IRS system show that SO emission is confined to the disk surface layers \citep{vanthoff2023}, which is at odds with the `centrifugal-barrier' model that would predict the SO emission peak to be within the disk midplane.

The specifics of the model adopted to describe the transition from envelope to disk aside,
there is likely some shock and/or transition that affects the chemical makeup of the
gas-phase molecules initially within the disk. This is evident in the molecular transitions
observed in L1527 IRS with carbon chain molecules and SO \citep{Sakai2014a, Sakai2014}, 
some of the observed SO rings \citep{Yen2014}, and the lack of \hcop\ strongly tracing
some of the youngest disks \citep{Reynolds2021}.

\section{Protostellar Disks}

\begin{figure}
    \centering
    \includegraphics[width=5in]{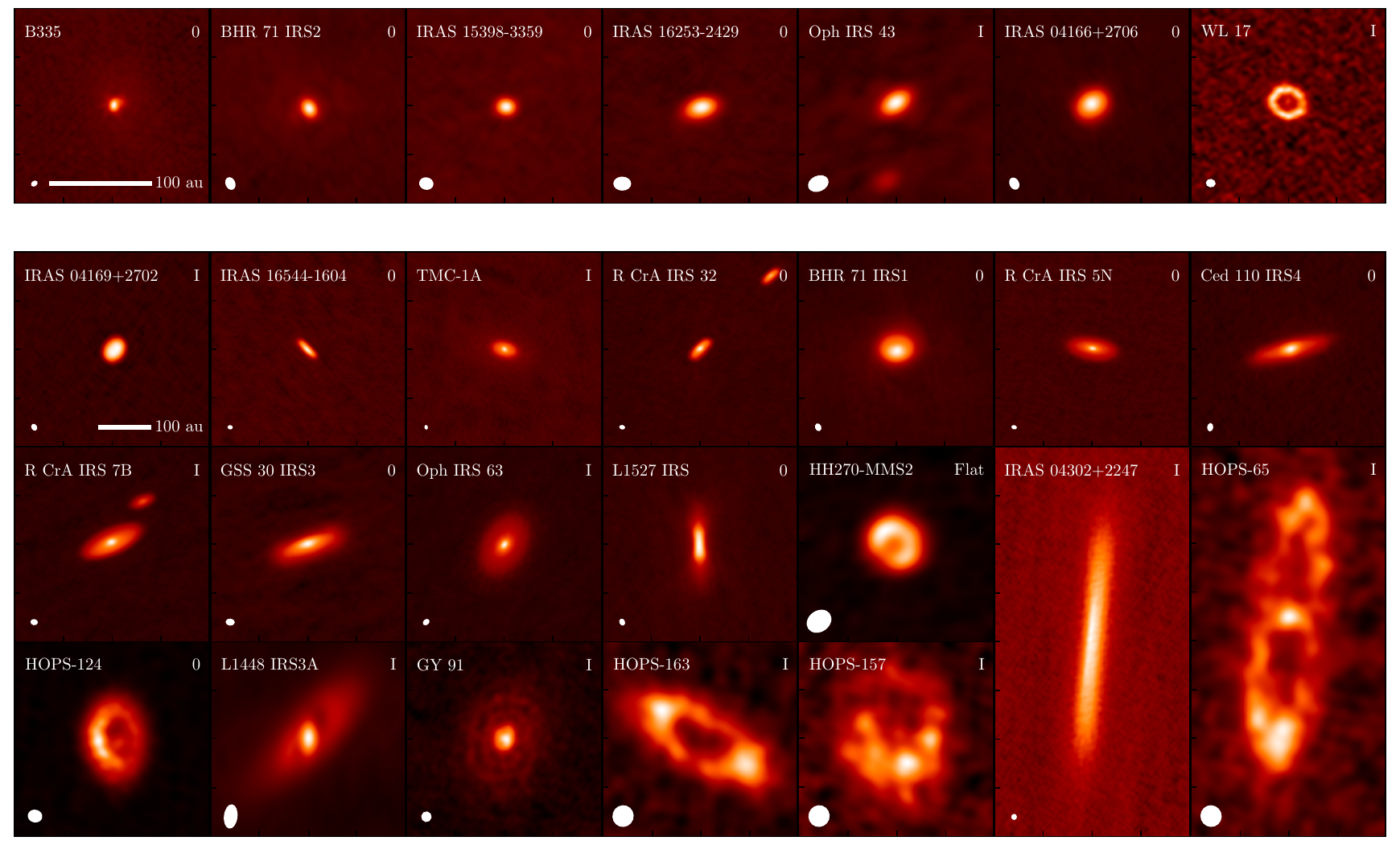}
    \caption{Gallery of protostellar disks demonstrating a broad range of disk structures including a range of sizes, brightnesses, and substructures. We note that the first row is zoomed in by a factor of two in physical size to better showcase the structures of small disks, while the remainder of the rows have the same physical size to demonstrate the broad range of protostellar disk sizes. Data are from \citet{Tobin2020}, \citet{Ohashi2023}, \citet{Segura-Cox2020}, \citet{Reynolds2021}, \citet{Sheehan2017a}, \citet{Sheehan2018}, and \citet{Sheehan2020}.}
    \label{fig:disks_gallery}
\end{figure}

Investigations over the past two decades have conclusively demonstrated the existence of rotationally-supported disks around Class 0 protostars as well as around those in later evolutionary classes (e.g., Figure \ref{fig:disks_gallery}). Protostellar disks form as a direct consequence of angular momentum conservation during collapse and mark the initial conditions for the planet formation process. As such, we will review the basic methods for characterizing the disks, the characteristics of ensemble populations, and the disk physical structure.

\subsection{Methods for Analyzing Disk and Envelope Structure}
\label{section:disk_methods}
Studies of protostellar disk and envelope structures are made difficult by the complicated geometry of the systems. Indeed at a range of spatial scales emission from disk and envelope are entangled, and studying the properties of either necessitates techniques for separating disk and envelope.

Prior to ALMA, spatially resolving protostellar disks was only feasible for a few of the nearest and largest disks \citep[e.g.,][]{Wolf2008,Tobin2012}, though some of the exceptional cases viewed in scattered light are noteworthy \citep{Padgett1999}. Early studies of protostellar disks leveraged the power of the $uv$-plane to identify disk-like features towards protostars by looking for a flattening of visibility profiles towards the longest baselines in interferometric observations. This feature indicated the presence of an unresolved, compact structure likely to be the disk. The baseline at which the visibilities flattened and over which this flattening extended to could be used to place constraints on the size of said unresolved feature \citep[e.g.,][]{Harvey2003,Looney2003,Chiang2012}. Similarly, as an extension of this method it was postulated that by picking a baseline thought to represent the size scales of disks, one could use the flux density on that baseline to estimate the disk flux density, and thereby the mass of the disk, using the standard flux-to-mass conversion \citep[e.g.,][]{Hildebrand1983}. A survey by \citep{Jorgensen2009} adopted this method to estimate the masses of a collection of protostellar disks, as have some more recent studies with limited resolution \citep[e.g.,][]{Andersen2019}. Simulated observations by \citet{Dunham2014}, however, found that the masses estimated by such tools can be significantly incorrect, suggesting that masses derived in this way should be treated with caution.

Alternatively, forward modeling using radiative transfer codes to fit interferometric (among other) observations has also provided a way to study the structures of the environments around protostars. In this paradigm, a parametric model describing the protostar, disk, and envelope is supplied to a radiative transfer code that is then used to generate synthetic observations of that system. Those synthetic observations are then compared with actual observations, tuning parameters to find a set that fit. Due to the computational expense of generating radiative transfer models, the scope of such models was initially limited to small numbers of sources \citep[e.g.,][]{Harvey2003,Chiang2012}. Large grids of models that varied parameters in a regular manner enabled such models to expand to larger sample sizes \citep[e.g.,][]{Robitaille2006,Eisner2012,Furlan2016}. Such grids, however, were necessarily coarse and so the best-fit parameters were often sampled poorly and uncertainties were difficult to assess. In recent years, studies have begun to leverage the computational resources made available by supercomputers to employ more sophisticated fitting tools such as Markov Chain Monte Carlo (MCMC) to rigorously fit radiative transfer models to multiple independent datasets \citep[e.g.,][]{Sheehan2022a}, enabling best-fit parameter values and uncertainties to be measured with greater fidelity. This technique is, however, limited by systematics in the model choice.

In many ways ALMA settled the debates with regard to the existence and properties of protostellar disks. The high angular resolution afforded by ALMA yields images of protostellar systems with enough resolution to directly resolve the size scales of disks and thereby study the geometry of these systems directly. Such studies have revealed disk-like emission on $\lesssim$100~au scales that are typically associated with Keplerian rotation when observed with appropriate spectral lines. Furthermore, in the newly emerging high resolution observations with ALMA, clear disk structures are seen \citep[Figure \ref{fig:disks_gallery}; e.g.,][]{Sheehan2017a,Sheehan2018,Segura-Cox2020,Ohashi2023}. Modeling shows that the disk represents a large density increase from the envelope \citep{Aso2017}, such that mistaking the infalling envelope for the disk in continuum observations is unlikely. Furthermore, whenever a disk has been clearly resolved in the dust continuum it is usually accompanied by molecular emission that can be characterized as Keplerian rotation \citep[e.g.,][]{Tobin2012,Murillo2013,Ohashi2014,Harsono2014,Yen2017,Maret2020,Reynolds2021,Ohashi2023}. Comparison of disk flux densities derived from radiative transfer forward modeling with flux densities derived from Gaussian fitting at moderate resolution for large samples of sources has demonstrated that simple tools can be good proxies for quantities like disk flux densities. This suggests that these compact features are indeed associated with protostellar disks. While high resolution images can therefore be used to study disk structures, confounding factors such as optical depth, opacity values, viewing angle effects, etc. limit the utility of such simple metrics for studying disk structure. Future studies may require radiative transfer modeling or some in-between to disentangle these effects from the underlying physical structure in order to answer more challenging questions, such as the state of grain growth in young disks or the true underlying disk mass distribution, to name a few.

\subsection{Disk Masses}

Disk masses have traditionally been estimated by assuming that disks are optically thin, isothermal, and have a constant dust opacity throughout. The dust masses can then be calculated from the millimeter flux density via
\begin{equation}
    M_d = \frac{F_{\nu} \, d^2}{B_{\nu}(T) \, \kappa_{\nu}},
    \label{eq:flux-to-mass}
\end{equation}
where $B_{\nu}$ is the Planck function, $F_{\nu}$ is the disk flux density, $T$ is the disk temperature, $d$ is the source distance, and $\kappa_{\nu}$ is the dust opacity at the frequency of the observation. From there, if total disk mass is desired then one traditionally multiplies the dust mass by a gas-to-dust ratio of 100 to get the total mass of the disk.

As discussed in the previous section, the compact dust emission on $<$100 au scales can now confidently be associated with protostellar disks, so the disk demographics studies that have been carried out toward Class II disks in regions of different ages \citep[e.g.,][]{Ansdell2016,Pascucci2016} can be extended to protostars without absolutely requiring detailed modeling of the disk and envelope physical structure \citep{Sheehan2022a}. 
However, the conversion of flux density to mass is confused by the need to adopt a dust opacity law that provides consistent masses for observations at different frequencies, the average dust temperature and its variations with disk radius and system luminosity, and the optical depth of the dust emission itself \citep[e.g.,][]{Tobin2020,Sheehan2022a}.
As such, it is prudent to establish what is known observationally by only comparing datasets taken at the same observing frequency separate from a discussion of what may be inferred about disk masses, which can be heavily influenced by the assumptions made in utilizing Equation \ref{eq:flux-to-mass}.

\begin{figure}
    \includegraphics[width=5in]{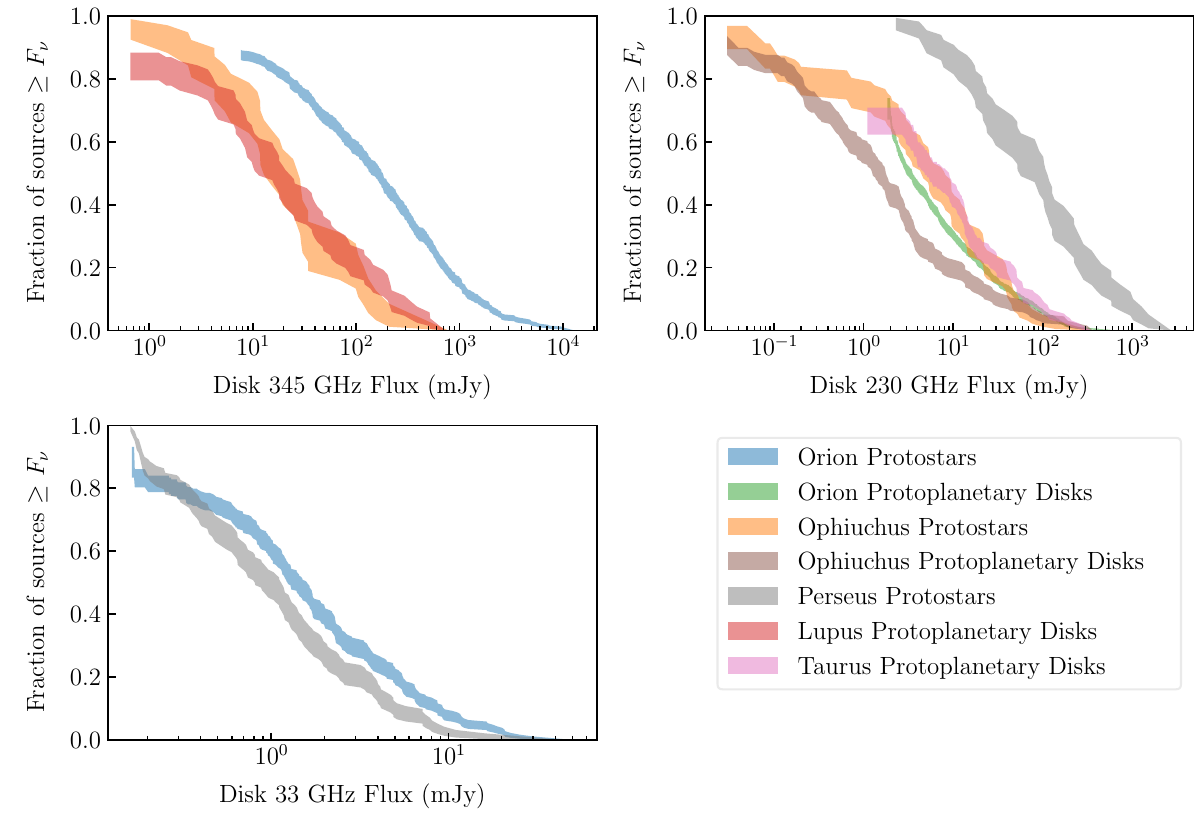}
    \caption{Cumulative distributions of disk flux densities, scaled to a common distance of 140 pc, for an assortment of different star forming regions and disk ages (i.e., protostar vs. protoplanetary disk) at the three frequencies where such studies of protostellar disks have been most common. We note that although we include Perseus protostars in the 230 GHz panel, much of the data used in that survey was actually taken at $\sim265$ GHz and therefore these disks may be somewhat more luminous than if they were truly 230 GHz flux densities, though we do not expect this difference to affect our conclusions.}
    \label{fig:flux_distributions}
\end{figure}

In that vein, we have collected the disk flux densities measured for protostars from surveys of protostellar disk properties in Orion \citep{Tobin2020}, Perseus \citep{Tychoniec2018,Tychoniec2020}, and Ophiuchus \citep{Cieza2019,Williams2019,Encalada2021}. These surveys were taken across three separate observing frequencies (345 GHz -- Orion and Ophiuchus; 230 GHz -- Ophiuchus and Perseus; 33 GHz -- Orion and Perseus), and critically we only compare observations taken at very similar observing frequencies. We also collect the surveys of protoplanetary disks carried out at 345 GHz and 230 GHz \citep[e.g.,][]{Andrews2013,Ansdell2016,vanTerwisga2022} for comparison with respective flux densities for protostellar disks, since the time evolution of disk flux densities is often as of much interest as comparison across regions.
We show cumulative distributions of disk flux densities, scaled to a common distance of 140 pc, for each region in Figure \ref{fig:flux_distributions}.

While the inhomogeneous set of protostellar disk survey frequencies makes it difficult to compare all of the samples in a uniform way, the VLA observations of Orion and Perseus at 33~GHz suggest that the luminosity of disks in these two regions is comparable. Conversely, the 345 GHz flux densities of Ophiuchus protostellar disks are significantly lower (on-average) than those of Orion protostars and their 230 GHz fluxes are also below those of Perseus, so it is not necessarily the case that protostellar disks in all regions are similar. Ophiuchus, however, has significant foreground extinction that makes classification of the protostars notoriously difficult and classifications do not always agree \citep{Enoch2009,VanKempen2009, Mcclure2010}.
It can also be seen from Figure \ref{fig:flux_distributions} that Orion protostars are, on average, more luminous at 345 GHz than protoplanetary disks in Lupus and the Perseus protostars are on average more luminous at 230 GHz than protoplanetary disks in Taurus and Orion. Taurus, Lupus, and Orion appear generally representative of the protoplanetary disk populations. We note that the difference in wavelength between the Orion protostar and protoplanetary disk surveys prevents a direct comparison. However, the Orion and Perseus protostars have comparable luminosities at 33 GHz and Perseus protostellar disks are more luminous than Orion protoplanetary disks at 230 GHz. Taken together, this evidence strongly suggests that the Orion protostellar disks are more luminous than the protoplanetary disks, though direct confirmation of this is needed. Although Ophiuchus protostars are similar in luminosity to typical protoplanetary disks at both 230 GHz and 345 GHz, it is worth noting that they do follow the same trend when compared to protoplanetary disks {\it within the same region}. So there is reason to believe that protostellar disks tend to be more luminous at sub/millimeter wavelengths, on-average, than protoplanetary disks with variations on the median disk luminosity from region to region.

With these observational trends established, the question of how disk {\it masses} vary between regions and evolution becomes murkier. A straightforward application of Equation \ref{eq:flux-to-mass}, or even one that attempts to scale the temperature with protostellar luminosity \citep[e.g.,][]{Tobin2020},  more or less preserves the trends seen in Figure \ref{fig:flux_distributions}. Indeed, when comparing the disk masses inferred from Orion 345 GHz flux densities, \citet{Tobin2020} found that protostellar disks are only 2.5 -- 8$\times$ more massive than the typical protoplanetary disk population, and \citet{Williams2019} found that Ophiuchus Class Is are $\sim5\times$ more massive than the Class II population. Moreover, further work by \citet{Tychoniec2020} showed that disk masses for Perseus measured using a collection of ALMA data at 230 and 265~GHz were $3 - 10\times$ more massive than protoplanetary disks in Lupus, and \citet{vanTerwisga2022} found that protostellar disks in Orion are a factor of $13 \pm 3$ $\times$ more massive than protoplanetary disks. It should be noted, though, that both of these comparisons crossed observing wavelengths. That said, the differences seen for the Ophiuchus protostars provides evidence that protostellar disk masses could vary from region to region (if classification is not a severe problem in Ophiuchus).

In an alternate approach, \citet{Sheehan2022a} recently published disk mass distributions for a subset of Orion protostellar disks
that were derived from forward modeling of the disk and envelope emission using radiative transfer models of 97 systems. Notably, the
dust disk masses they derived are substantially less massive than those found from a simple application of Equation \ref{eq:flux-to-mass}. This difference originates from the more realistic treatment of temperature in the radiative transfer models and
the overall larger dust mass opacity used in the models. As the disks in the sample tended to be small, the net effect was to increase the temperature and thereby decrease the mass. This is consistent with previous studies showing that young disks tend to be warmer \citep[e.g.,][]{vantHoff2018UnveilingL1527,vantHoff2020TemperatureWarm} due to the envelope keeping radiation from escaping and heating the outer regions of the disk. 
In fact, when comparing the disk masses for Class II disks derived from similar radiative transfer forward modeling by \citet{Ballering2019ProtoplanetaryTaurus}, they found that protostellar disks were actually {\it less} massive than protoplanetary disks, in contrast with what is inferred from the disk flux densities.

There are caveats to these results, including differences in regions sampled and the differing maximum dust grain sizes from the fits, which is degenerate with dust mass (though the differing maximum dust grain sizes between the protostellar and protoplanetary disks could also demonstrate physical evolution with time), so ultimately it remains unclear how masses compare between protostellar and protoplanetary disks. Regardless, these studies demonstrate the difficulty and ambiguity in deriving disk masses from observational properties such as flux densities, and highlight the need for careful, consistent, and uniform modeling that can accurately and simultaneously constrain dust properties alongside disk properties while accounting for temperature and radiative transfer effects in a uniform way across the various samples of disks.

\subsection{Disk Radii}

As was discussed in Section \ref{section:disk_methods}, the angular resolution afforded by ALMA has enabled protostellar disk radii to be measured with minimal confusion from the envelope.
Dust disk radii from millimeter images are the most straightforward to measure because the infalling envelope has very low optical depth and is largely filtered out
in observations with high enough resolution to resolve the disks. That said, there are few systematic surveys with enough resolution to well-resolve protostellar disks at this time, and other studies with more limited resolution are difficult to compare fairly. 
It is also important to point out
that methods of measuring dust disk radii vary between different studies, with some using, for example the radius containing 90\% of flux while others use Gaussian widths or even sophisticated functions to fit the data. Thus, a meta study of disk radii from the protostar literature to-date can be misleading as systematic differences between studies will be confounding to the distribution \citep[e.g.,][]{Tsukamoto2022}.

It should also be noted that the dust in disks may experience additional dynamical effects that could impact relationships derived using dust disk radii. In particular, due to the decreasing density (and therefore pressure) of a disk as a function of radius, the gas in a disk is expected to be at a slightly sub-Keplerian rotation speed. This means that the dust feels a headwind as it orbits in the disk, and this headwind removes angular momentum from the dust, causing it to drift radially inwards \citep[e.g.,][]{Weidenschilling1977AerodynamicsNebula}. As such, the growth of the {\it gas} disk could be counteracted by the inward drift of solids in the disk. Indeed observations of gas disk radii in protoplanetary disks have found they are typically a factor of a few larger than measured dust disk radii, though it remains unclear whether this is broadly due to radial drift or line versus continuum optical depth effects \citep[e.g.,][]{Trapman2019}. Even gas disk radii may not be immune to additional effects. Recent work suggests that disks may also lose angular momentum to disk winds \citep[e.g.,][]{Bai2013}, also counteracting this growth. With this in mind, we focus on what is known about protostellar dust disk radii throughout this section, and return to gas disk radii in the next section.

\begin{figure}
    \centering
    \includegraphics[width=5in]{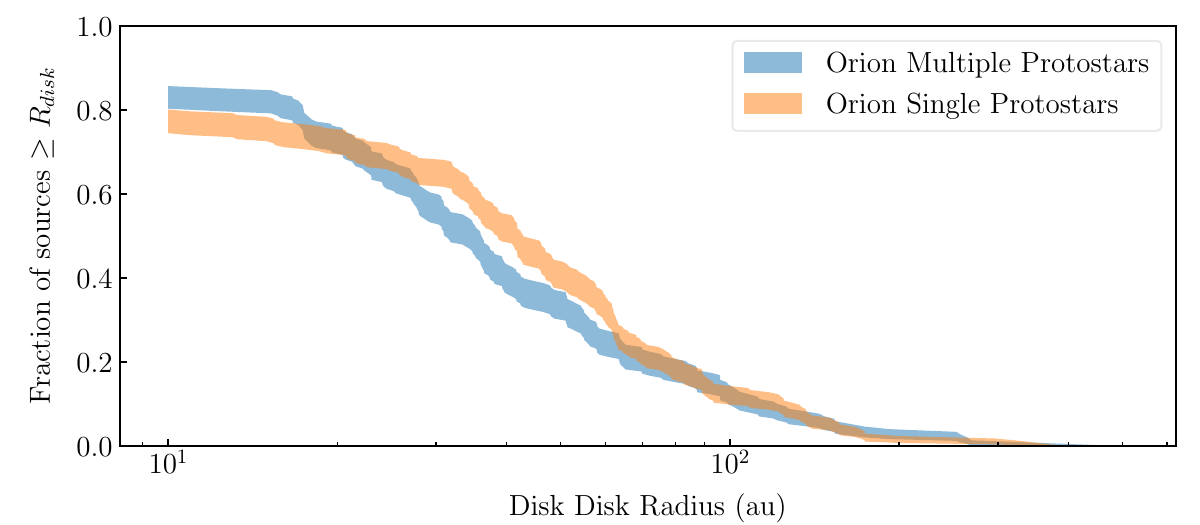}
    \caption{Cumulative distributions of dust disk radii, as measured by the deconvolved major axis of a two-dimensional Gaussian fit to each source, for protostars in Orion, split between single protostars and multiples.}
    \label{fig:disk-radii}
\end{figure}

The most homogeneous survey of protostellar disk dust radii was performed in \citet{Tobin2020} toward protostars in Orion. That survey used the 2$\sigma$ radii of
Gaussian fits to measure the disk radii of the full sample from the dust continuum images. Their results showed that protostellar disks have median radii $\lesssim$50 au for protostars of all classes. We show the cumulative distribution of disk radii from this survey in Figure \ref{fig:disk-radii}. 
While the median radii for Class 0 disks are larger than those for Class I and Flat Spectrum protostars, the statistical evidence for true differences in the radii distributions between classes is weak \citep{Tobin2020, Sheehan2022a}. Moreover, there is scatter in disk radii over an order of magnitude within each protostellar class, contradicting expectations of disk growth with evolution, in so far as the classes trace evolution. 

Figure \ref{fig:disk-radii} also shows the disk radii split into single systems and multiples for the Orion protostars. From these distributions, without subdividing the sample in terms of companion separation, there is not statistically significant evidence for the samples to be drawn from different distributions. However, when splitting the sample more finely in terms of companion separations, \citet{Offner2022} found that the disks with companions separated by $>$300~au were larger than the disks for multiple systems with companions at $<$ 300~au separation. The distribution of disk radii for non-multiple systems  was in between those two samples.

The measured disk radii do provide important constraints on models of disk formation, for which disk radius is an important discriminator. 
The observations of protostellar disk radii are at odds with the expectation from models of disk formation that disk radii should increase with time during protostellar collapse \citep{Terebey1984, Ulrich1976}, as higher angular momentum material reaches the disk. The fact that disk radii do not change, as a population, may not be 
surprising, however, because the increase of disk radius with evolution applies to a single system, while
a population of protostars is combining measurements of disk radii from protostar systems that likely had a wide variety of initial conditions.

Hydrodynamic models, with conserved angular momentum
from an infalling envelope with initially solid-body rotation \citep{Yorke1999, Vorobyov2011}, have tended to produce disks that are several hundred au 
in radii and also massive. On the other hand, MHD models (with non-ideal MHD) typically produced disks that were $\sim$10s of au or
less \citep{Wurster2019a,Hennebelle2016}, and even earlier models using only ideal MHD could completely suppress disk formation \citep{Allen2003}. 
The disk radii derived from global hydrodynamic simulations \citep[without magnetic fields;][]{Bate2018} tend to produce disk radii that are larger than predictions of MHD, but smaller than the aforementioned single source simulations without magnetic fields. The fact that global simulations do not produce excessively large disk radii could stem from the fact that cores get angular momentum from gravitational torques rather than initial rotation of the cloud \citep{Kuznetsova2019}.

The relatively small protostellar disk radii measured could be interpreted as evidence that the effects of magnetic fields are important for regulating disk formation. \citet{Tobin2020} did find that the dust disk radii in Orion were comparable to the gas disk radii from the \citet{Bate2018} hydrodynamic simulations. But, differences in disk radius measurement techniques between the models and observations (the models did not use simulated observations), and the systematic overestimates of the small disk radii from \citet{Tobin2020} as pointed out by \citet{Sheehan2022a}, drives some further disagreement between the hydrodynamic models and the observations. Thus, the large spread of possible initial angular momenta and magnetic field strengths may result in the observed distribution of disk radii.

The recent higher resolution survey by \citep{Ohashi2023} further illustrates the distribution of disk radii in the protostellar phase. For systems where the disks
would have been unresolved or marginally resolved previously, they clearly resolve small dust disks that also appear to be in Keplerian rotation. 
Thus, there appears to be a continuum of protostellar disk sizes from large to small disks. It remains unclear, however, what is happening with the smallest
apparent disks where only hints of resolved structure are apparent at $\sim$5~au resolution \citep{Bjerkeli2023,Ohashi2023}. These could be cases of either strong magnetic fields that are dynamically important, or the systems are just very young and the disk has not grown in size.

\subsection{Gas Disk Properties}

Measurements of gas disk radii and gas disk masses are faced with significantly more complexities. The molecular line emission from the envelope near the disk can be quite significant, confusing an observational determination of the transition from envelope to disk (e.g., see Figure \ref{L1527_kinematics}). Moreover, the high optical depth of the CO isotopologues in protostellar disks \citep[even \cateo\ can be optically thick,][]{vantHoff2018} makes the approach used in dust disk observations, where the brightness of the disk is significantly higher than the extended envelope, ineffective. As such, systematic characterizations of the gas disk masses relative to the dust disk masses that have been performed on protoplanetary disks \citep[e.g.,][]{Ansdell2016} are not presently available for protostellar disks. There is potential for using \csevo\ to characterize protostellar gas disks since it is more optically thin and tends to have less confusion with the envelope \citep[e.g.,][]{vantHoff2020TemperatureWarm,Reynolds2021}.

Thus, the primary method for measuring gas disk radii has depended on kinematic measurements to separate the expected Keplerian disk from the infalling/rotating envelope. Studies of gas disk radii to date have either relied on fitting the rotation curves from the observed PV diagrams \citep[e.g.,][]{Yen2013,Yen2017}, fitting the marginally-resolved emission channel-by-channel and plotting the observed velocity as a function of fitted position \citep[e.g.,][]{Tobin2012,Murillo2013,Harsono2014}, or forward modeling of the molecular line emission of both the disk and envelope \citep[e.g.,][]{Tobin2020,Reynolds2021}. In the two former approaches, without radiative transfer modeling, the radius of the Keplerian (or gas disk) is defined by the measured change in rotation curve slope from v~$\propto$R$^{-0.5}$ to v~$\propto$R$^{\sim-1}$.

\citet{Yen2017} published a compilation of protostellar gas disk radii. They attempted to infer trends in gas disk radii, protostar mass, and mass accretion evolution. The results can fit some simple models of disk radius evolution with protostellar class (and inferred accretion age), and they find some evidence of a linear correlation between protostar mass and disk radius. However, the small numbers and bias in the sample of masses toward solar-mass protostars make firm conclusions about the correlation of disk radius and protostar mass uncertain. Much like the meta analyses of disk radii referenced earlier \citep{Tsukamoto2022}, the heterogeneous nature of the measurements and the sample selection are likely to have systematic errors relative to each other. Thus, future studies would benefit from a uniform approach to measuring gas disk radii that can produce reliable uncertainties (or are at least internally consistent). The gas disk radii from the \citet{Ohashi2023} survey also do not yet have a uniform analysis, which would help to better understand the relationship between the gas and dust disks given the uniformly observed sample.

\subsection{Protostellar Disk Sub-structure}

One of the main highlights of the first five years of ALMA was the discovery of bright rings and dark gaps, collectively termed ``substructures" in the disk around the pre-main sequence star HL Tau \citep{Brogan2015}. These features were initially, and excitingly tied to the presence of planets hiding within the disk, carving the material. Although, numerous other explanations ranging from chemistry, dust optical properties, and magnetic properties of the disk have been proposed to explain such features without the presence of planets. Regardless of whether planets are indeed the underlying cause of these features, it is almost certainly the case that these features are signposts of ongoing planet formation as regardless of their origin, they should be locations where the potential for planet formation is enhanced. Moreover, systematic studies of substructures in protoplanetary disks (ages $\gtrsim$ 1~Myr) have emphasized that these features appear to be ubiquitous, at least within disks that are large enough to resolve such features \citep[e.g.,][and references therein]{Andrews2020,Bae2022}.

The prevalence of said features in $\gtrsim1$ Myr-old disks naturally raised the question of how early these features form, and as a corollary, how early planet formation may begin. Early signs suggested that such features might be similarly common in young disks: HL Tau itself is typically classified as a Flat Spectrum protostar \citep{Furlan2008} indicating that it may have some amount of remnant envelope material. \citet{Sheehan2017a} found the first case of an embedded ``transition disk", with a single large cavity in the center, while \citet{Sheehan2018} found the first embedded disk analogous to HL Tau in GY 91, with three gaps within its large disk. Additional rings, gaps, and asymmetries have subsequently been found in a handful of additional disks \citep{Ohashi2022,Yamato2023,Sheehan2020,Sheehan2022b}, including in the disks of the well known protostars L1489 IRS and L1527 IRS.

Aside from the numerous protostellar disks found with large cavities and/or asymmetries, many of the protostellar disks with substructures tended to be classified as Class I or Flat Spectrum. This could mean that perhaps there was little envelope material left and that these sources were not all that dissimilar in age to the protoplanetary disks observed with substructures. One of the more intriguing disks with substructures found was Oph IRS 63, as it has served as a preview of what was to come. \citet{Segura-Cox2020} found that the disk of Oph IRS 63 has two bright rings above the otherwise smooth disk background; however, these features are faint enough that they are not easy to pick out by-eye in the way that many of the previously described substructures were. A survey of 10 protostellar disks at high angular resolution found either weak features or no features, suggesting that prominent substructures may not often form by the end of the embedded phase \citep{Cieza2021TheResolution}.

This ultimately leads to the recent  systematic survey of 19 protostellar disks at $\sim$0\farcs04 angular resolution with ALMA. Intriguingly, and in-line with \citet{Cieza2021TheResolution}, they found only a limited number of substructures in the disks surveyed \citep[e.g.,][]{Ohashi2023}. Though a more detailed characterization of weak substructures is still forthcoming, the conclusion appears to be that substructures have not yet formed or are weak in amplitude early in the protostellar phase, but that they develop and become more detectable late in the protostellar phase. Planet formation could still occur or begin in the protostellar phase, however, as substructure formation is not the first step of this journey, but it certainly seems that the physics responsible for generating substructures has not had sufficient time in this stage to develop into strong, prominent features in dust continuum images.

\subsection{Dust Opacities, Grain Growth, and Settling}

The lack of ubiquity of substructures in protostellar disks, as compared with older disks, suggests that massive planets have not yet formed, but this does not mean that planet formation has not progressed. Indeed the first step in the planet formation process is the settling of dust grains to the midplane, and their subsequent growth into millimeter or centimeter-sized ``pebbles", which then can efficiently collapse into planetesimals via the streaming instability \citep[e.g.][]{Simon2016}, or can be fodder for planetary embryos via pebble accretion \citep[e.g.][]{Lambrechts2012}. As such, a natural question to ask is whether there is any evidence for dust grain growth protostellar envelopes or disks

Some attempts have been made to address this question on different scales. On the scale of the protostellar core, there are indications of at least $\sim$1~\micron-sized dust grains in the outer parts of the cores from observations of diffuse scattered light at 3.6 and 4.5~\micron, a phenomenon dubbed `coreshine' \citep{Lefevre2014}. Constraints on the dust grain sizes in the inner envelopes, however, requires interferometry at submillimeter/millimeter wavelengths. Early studies of the inner envelopes around protostars found evidence, via measurements of the millimeter spectral index, that some dust grain growth had occurred in the inner envelope. \citet{Kwon2009} found a typical value of $\beta = \alpha - 2$, where $\alpha$ is the millimeter spectral index, of $\sim1$ for three protostars \citep[compared to $\beta$ $\sim$1.7 for small dust grains,][]{Ossenkopf1994}, suggesting some grain growth may have happened in the envelope, though the lack of a disk component in their model may have affected their results. Further work by \citet{Miotello2014} and \citet{Cacciapuoti2023} have provided additional claims of dust grain growth in the inner envelopes of protostars, while conversely, \citet{Agurto-Gangas2019RevealingPer-emb-50} found no or limited evidence of dust grain growth in the envelope of a protostar in Perseus.

Among the more compelling claims of evidence for dust grain growth in protostellar disks is a study of TMC-1A by \citet{Harsono2018}. In that work, the authors found that the C$^{18}$O emission showed a large central cavity. They suggest that rather than originating from a true deficit of C$^{18}$O, this cavity is the result of the emission being hidden behind optically thick dust, giving it the appearance of a cavity. Through their modeling they suggest that in order to achieve the opacities needed to reproduce such shielding, dust grain growth to millimeter-sizes is necessary. Outside of this work, however, there has been limited effort to measure dust grain growth in young disks, likely stemming from the difficulty of separating the disk from the envelope.

Though measurements of dust grain sizes have been somewhat limited, interestingly and conversely, a number of recent studies {\it have} found that for a few Class 0/I sources, the vertical extent of the disk is still quite significant, indicating that dust grains have yet to settle to the midplane \citep[e.g.,][]{Sheehan2022b,Lin2023,Ohashi2023,vanthoff2023,Villenave2023}. As settling to the midplane is a precursor to even the dust grain growth process from micron to millimeter sizes, the lack of settled dust in these young disks may provide a lower bound of the timescale for planet formation. If so, then the earliest disks may be dominated by the settling of dust to the midplane, with dust grain growth taking hold late in the protostellar phase.

\subsection{Protostellar Disks and Multiplicity}

The aforementioned surveys of protostellar disks in nearby star forming regions have also revealed that 30\% of protostars form with another YSO within 10,000~au in projected separation \citep{Tobin2022} and $\sim$60\% of these multiple systems are expected to evolve together as coeval systems \citep{Murillo2016}, see Figure \ref{fig:disks_gallery} for a few examples. Moreover, the distribution of separations appears double-peaked with peaks at $\sim$100~au and $\sim$3000~au \citep{Tobin2022}. Core fragmentation is the obvious mechanism to produce companions at $\sim$1000~au separations, but the origin of the $\lesssim$~500~au companions is less clear. Both core fragmentation with inward migration and gravitational instability of the protostellar disk are possible \citep[see review by][]{Offner2022}, and consensus about the relative contributions from each mechanism is still lacking.

Whether or not protostellar disks often go through a gravitationally unstable period is an important consideration for the feasibility of disk fragmentation.
\citet{Tobin2020} estimated that only 6 out of 289 detected protostellar disks were likely to be gravitationally unstable by assuming that the total disk mass was equivalent to $\sim$100$\times$ the dust mass. However, some of the stable disks were already part of a $<$~500~au multiple system. 
The small number of possibly unstable disks seems consistent with the lack of detected spirals from very high spatial resolution surveys \citep{Ohashi2023}, a tell-tale sign of gravitational instability. Some known spirals appear to be related to companions within disks \citep[e.g.,][]{Reynolds2021}, but in systems without companions, they could also result from the infall of the envelope to the disk \citep{Lee2020,Kuznetsova2022}.

\section{Protostar Masses}
The masses of protostars themselves remain one of the most difficult quantities to characterize. The embedded nature of protostars prevents their spectral classifications from being measured. Then in cases where spectra can be taken toward the protostars \citep[e.g.,][]{Greene2018}, the coexistence of features that could arise from both the protostar and the hot, inner circumstellar disk make spectral types further uncertain. Thus far, the exquisite spectra from JWST have told us little about the masses of protostars \citep{Yang2022,Harsono2023}. Even if the spectral types of protostars could be reliably measured, the path to turn those effective temperatures and bolometric luminosities into stellar masses is unclear. The mass to radius relation depends on stellar birthline models, for which the stellar radius depends on whether accretion is cold or hot \citep{Hosokawa2011}. The distributions of bolometric luminosities may be related to stellar mass to some degree, but with the significant caveat that a majority of the luminosity comes from accretion (L$_{\rm total}$~=~L$_{\rm protostar}$~+~L$_{\rm accretion}$) and will vary \citep[e.g.,][]{Dunham2010}. Disentangling the relative luminosity contributions is difficult, especially without an independent measurement of accretion luminosity, but the measurement of protostar masses, coupled with stellar birthline models \citep[e.g.,][]{Hosokawa2011} can provide constraints on L$_{\rm protostar}$. Thus far, however, the lack of protostar masses has limited the study of protostellar evolution to metrics based on the SED and envelope.

\begin{figure}
\minipage{0.55\textwidth}
\includegraphics[width=\linewidth]{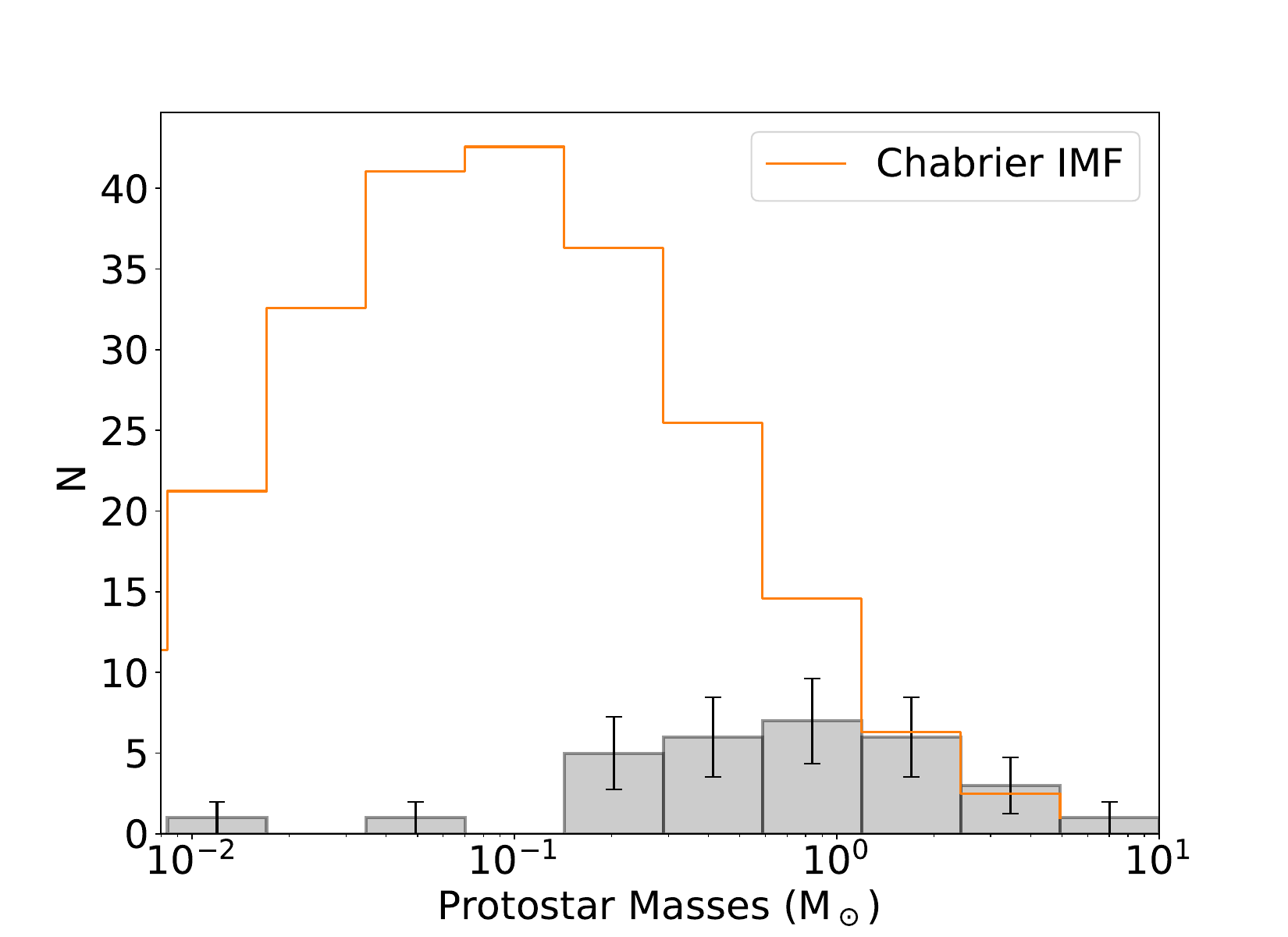}
\endminipage\hfill
\minipage{0.55\textwidth}
\includegraphics[width=\linewidth]{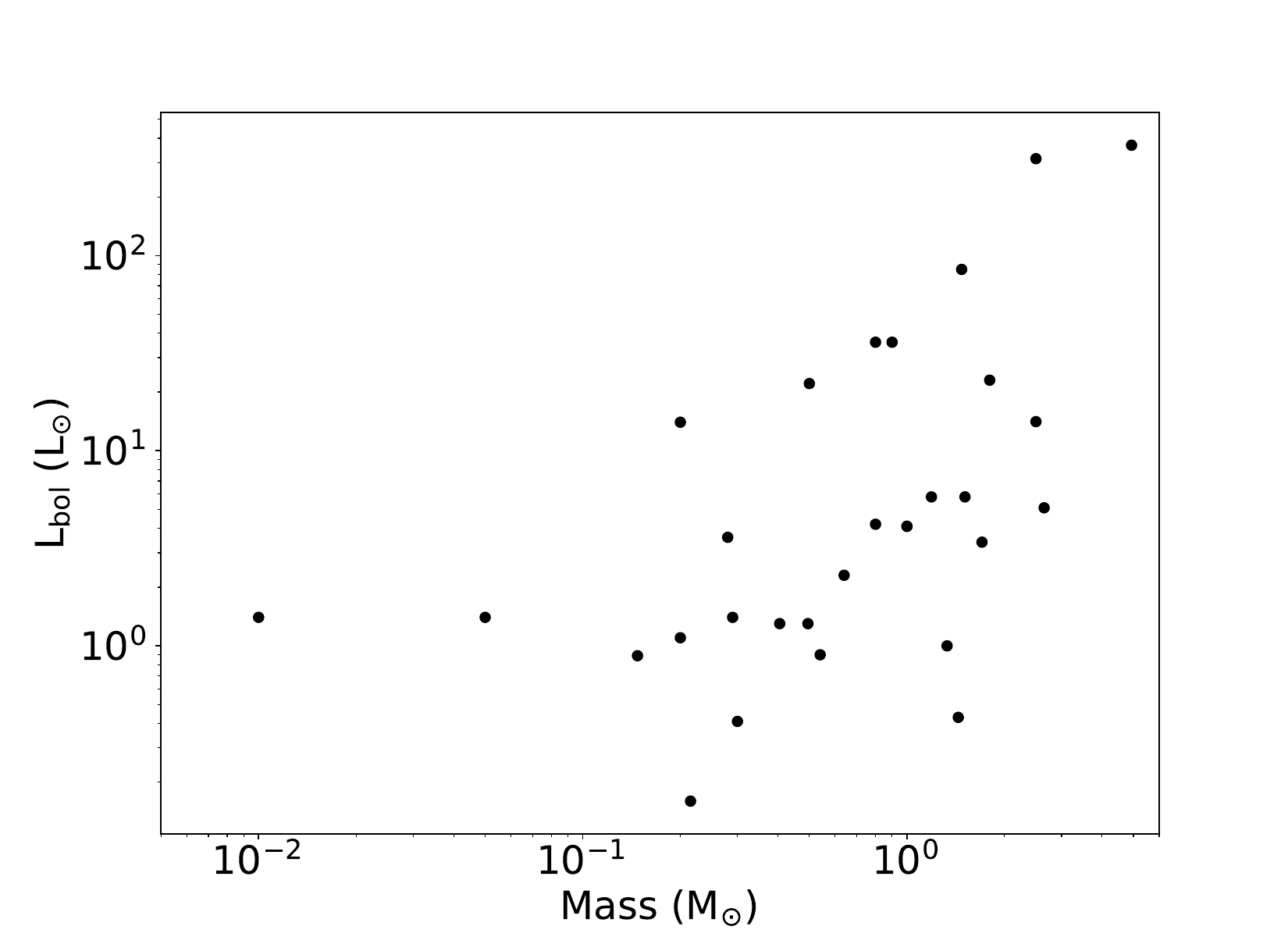}
\endminipage\hfill
\caption{Histogram of measured protostar masses summarized in Table \ref{tab1} with the \citet{Chabrier2003} initial mass function (IMF) overlaid and normalized to the highest mass with a populated bin (left). Plot of protostar mass versus \lbol\ (right). The IMF predicts that there should be significantly more low-mass protostars than are currently measured, demonstrating the bias in the sample of currently measured protostar masses.
}
\label{protostar_masses}
\end{figure}

The most successful methods for measuring protostar masses depend on molecular line observations at millimeter and submillimeter wavelengths to probe the Keplerian rotation of the disks surrounding the protostars and dynamically measure their masses. The alternative is to look for orbital motion of close binary protostars over many years \citep[to decades;][]{Maureira2020}. The Keplerian rotation method was pioneered on Class II disks using \twco\ observations \citep{Simon2000}, but could not readily be applied to protostars because \twco\ is optically thick and typically traces outflowing gas. Thus, isotopologues of CO or other less abundant molecular lines were required to trace the disks around protostars. Interferometers pre-ALMA had just enough sensitivity and resolution to characterize a few protostars \citep{Tobin2012,Harsono2014}, but ALMA and NOEMA (to a lesser extent) have enabled significantly more protostars to have their masses measured \citep{Murillo2013,Yen2017,Maret2020}.

There are a few complementary techniques used to derive protostar masses from molecular line data. One approach uses position-velocity (PV) diagrams, where one of the position axes is collapsed by averaging or summing the data along that direction (usually with a high aspect ratio). Then two approaches can be taken, fitting the peak of the emission in position and velocity space, colloquially known as the Ridge method \citep[e.g.,][]{Yen2013}, and fitting the maximum velocity as a function of position, known as the Edge method \citep[e.g.,][]{Seifried2016}. There is also a method that is similar to the Ridge method, where Gaussians are fit, channel-by-channel to the molecular line data to characterize the changing position with velocity; this fitting can be done on images or on the visibility data \citep{Tobin2012,Murillo2013,Harsono2014}. Then the fits to the data can explore a small range of parameters that include the power-law dependence of velocity and position, whether there is a change in the power-law at some radius (possibly signifying a transition from disk to envelope), and ultimately the protostellar mass.

The popularity of these data-driven methods stems from their speed and simplicity since they are not computationally intensive. However, there are drawbacks to using a 'reduced' set of data to characterize the masses of protostars. The Ridge method (and similarly the Gaussian fits per channel) picks the peak position of the emission as a function of velocity, but the orbital velocity (v=(G\mstar/R)$^{0.5}$) actually corresponds to the maximum velocity at a given radius, and when the molecular line emission from disk rotation is collapsed into a PV diagram, there is a superposition of multiple velocities along the line of sight. Thus, the Ridge method is expected to systematically underestimate the mass of the protostar, especially when the disk emission is well-resolved. The Edge method, which attempts to measure the maximum velocity along the line of sight also has systematic errors which result from the convolution of the beam and the data \citep[e.g.,][]{Maret2020}, which can push the fitted `edge' of emission to a larger radius at a given velocity. Higher-resolution data will have less of a bias, but the non-ideal nature of protostellar envelopes can create additional complications leading the Edge method to systematically overestimate protostar masses.

The alternative to these data driven approaches is to model molecular line emission and fit to either the full datacube or the visibility data. Approaches can vary from modeling the full radiative transfer \citep[e.g.,][]{Czekala2015,Tobin2020} to fitting isovelocity data scaled to more closely match the data \citep[e.g.,][]{Cheng2022}. These techniques have the advantage of fitting the full, three dimensional (position, position, velocity) dataset and leverage physically motivated models and often attempt to simulate the observations to disentangle the effects of the beam. However, the asymmetric nature of the envelopes around protostars (see Section 2) can cause difficulty for optimized fitting since the fitted models
are axisymmetric rotating, collapsing envelopes \citep[i.e.,][]{Ulrich1976} with embedded disks. Moreover, the disks themselves could also have asymmetries that are not reflected in the models \citep[e.g.,][]{Reynolds2021}.

Despite the caveats with regard to the different fitting techniques, the number of protostars with measured masses is now $\sim$30, and we have attempted to collect all of these masses in Table 1. Note that we do not include high-mass protostars in this sample, because the young stellar evolution of massive stars my be different from that of low to intermediate mass stars. We plot a histogram of these masses in Figure \ref{protostar_masses}, as well as the distribution of protostar masses compared with \lbol. The histogram immediately shows that the sample of measured protostar masses is biased toward solar-mass protostars, with a dearth of low-mass protostars compared with what would be expected from the initial mass function (IMF). The median mass provided in Table 1 is 0.8~\msun. Furthermore, the plot of protostar mass vs. \lbol\ shows that for each protostar mass, there are approximately 2 decades of scatter in luminosity, demonstrating the unreliability of \lbol\ as a mass proxy.

\section{Pathways for Improvement on Protostellar Classification}
\label{section:pathways}
\begin{figure}
     \includegraphics[width=5in]{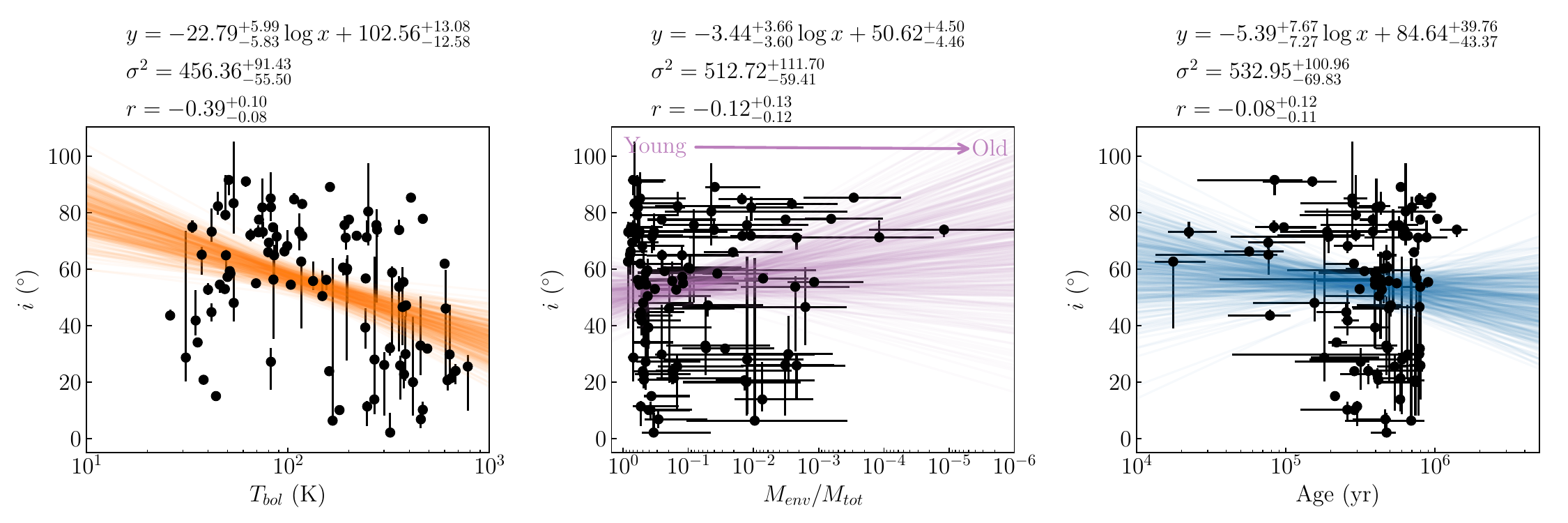}
     \caption{Comparison of various quantities that have been used/proposed as evolutionary tracers of protostars with inclination, with data taken from \citet{Sheehan2022a}. For each comparison, 100 samples from the posterior distribution of a linear fit to the data are shown as colored lines, and the best-fit values, variances, and correlation coefficients are listed. We see that $T_{bol}$, perhaps the most commonly used evolutionary tracer, is significantly correlated with inclination, while other tracers such as the ratio of envelope to total mass or the age estimated from simple evolutionary tracks are not. Adapted from \citet{Sheehan2022a}.}
     \label{fig:enter-label}
 \end{figure}

As discussed in Section \ref{section:classes}, it has been commonplace in the literature to use the protostellar classes, defined observationally, as a proxy for protostellar evolution. While this is motivated by the expected correlation between observational properties with physical properties as a system evolves from a deeply embedded protostar to a pre-main sequence star, it has also been established that the classes could be influenced by viewing orientation \citep[e.g.,][]{Crapsi2008,Furlan2016,Sheehan2022a}, see Figure \ref{fig:enter-label}. While viewing geometry is not the only property driving \tbol, its influence can confuse its connection to the evolution of protostar systems. Thus, as we move to a paradigm of more quantitative measurements of ensembles of disk properties and protostar masses, it is critical to find a metric that probes evolution without such pitfalls.

 An attempt to improve upon the Classes and replace them with physical Stages was made by \citet{Robitaille2006}. However, their definitions for protostars (Stages 0 and I) were based on ratios of envelope infall rate to stellar mass, and Stages II and III were based on envelope infall rates to stellar mass and disk mass to stellar mass. These definitions of Stages are problematic because envelope infall rate is notoriously difficult to measure \citep[e.g.,][]{Evans2015}, and in \citet{Robitaille2006} the infall rate is determined via SED fitting of radiative transfer models using rotating, infalling envelope models with embedded disks \citep{Ulrich1976,Terebey1984}. SED fitting alone is known to have many degeneracies, resulting in non-unique fits.

Alternatives probes of protostellar evolution also use the envelope as part of the evolutionary diagnostic, since this is a defining characteristic of protostar systems, but in a different way relative to \citet{Robitaille2006}. It is roughly expected that the importance of the envelope relative to the mass of other components of the system should decrease with time as the envelope is accreted onto the disk and star; however, such a relationship may not be monotonic.

Building on this and other similar ideas by \citet{Crapsi2008} and \citet{VanKempen2009}, \citet{Sheehan2022a} proposed using the ratio of envelope mass to the total mass of the system (including protostar mass) to characterize evolution. Though they used \menv\ fit from radiative transfer models and M$_*$ drawn randomly from the IMF (and included this as uncertainties on the measured values of \menv/M$_{\rm tot}$), envelope masses can be measured from sub-millimeter fluxes, and protostar masses are being measured more accurately. Most importantly, as demonstrated in Figure \ref{fig:enter-label}, this metric is not correlated with disk inclination.
Alternatively, \citet{Andre1994FromCloud} and subsequent works have also employed simple models of protostar and envelope evolution to produce protostellar evolutionary tracks in a \menv -- \lbol\ plane. By placing observed protostars in this \menv -- \lbol\ plane, ages for the systems can be estimated. Though these ages are inherently model dependent and not necessarily true ages, they do perhaps represent a way to estimate the relative evolution of an ensemble of protostars. This method is also not correlated with disk inclination (Figure \ref{fig:enter-label}), but it relies on \lbol, which is variable throughout the protostar phase \citep{Dunham2010} and not directly related to protostar mass.

Ultimately, any successful future classification system should be based on quantities that can be feasibly measured or constrained, are not strictly model-dependent, and 
are likely unidirectional metrics. By unidirectional metrics we mean that once a threshold is met a system would not revert to a previous state; the protostar mass is an example of a unidirectional metric. As such, the mass of the envelope relative to the protostar mas, and the mass of the envelope relative to the disk could be another route to determining the evolutionary state of a protostar, similar to the method proposed by \citet{Federman2023}.

\section{Searching for the Youngest Protostars}
Regardless of the classification system, none of the possible systems can uniquely identify the absolute youngest protostars.
The youngest protostars are of great interest because their current formation epoch, shortly after the collapse of their core, may reveal new insights into the physics of protostellar collapse and disk formation.
Some of the youngest protostars have the potential to be a first hydrostatic core (FHSC; note the different usage of the term `core'), where the compact central object ($\sim$several au in radius) has not yet risen to a high enough temperature to dissociate molecular hydrogen. Candidate FHSCs have been suggested from the observed properties of some systems \citep[e.g.,][]{Enoch2010, Pineda2011}, but unique characteristics of a FHSC that can be derived from observations are unclear.

Similar to (and sometimes overlapping with the candidate FHSCs) are the Very Low Luminosity Objects (VeLLOs). Candidate FHSCs and VeLLOs often have very faint near and mid-infrared emission and are surrounded by a dense core that often appeared starless prior to the advent of \textit{Spitzer}. These protostars are estimated to have internal luminosities $<$ 0.1 \lsun\ \citep{Young2004}, and are proposed as possible progenitors to brown dwarfs. However, many VeLLOs are surrounded by envelopes massive enough that, if even a few 10s of percent are accreted, they would be beyond the hydrogen burning limit. 

\begin{marginnote}[]
\entry{PACS}{Photodetector Array Camera and Spectrometer, an instrument on the \textit{Herschel Space Observatory} that operated between 60 and 210~\micron.}
\end{marginnote}
A distinct sample of very young protostars was discovered in Orion by the \textit{Herschel} Orion Protostar survey known as the PACS Bright Red Sources \citep[PBRS;][]{Stutz2013}. As their name suggests, these protostars have very red colors, very low bolometric temperatures, and were sometimes undetected shortward of 70~\micron. They are distinct from the VeLLOs and candidate FHSCs, because they have luminosities $>$1~\lsun. Further follow-up of the PBRS found that they have very bright, $<$1000~au radius emission at millimeter wavelengths, similar to some of the youngest and most millimeter bright protostars in the Perseus and Ophiuchus star forming regions \citep[e.g., NGC 1333 IRAS4A and IRAS 16293-2422;][]{Tobin2015b}. Resolved observations of the PBRS at submillimeter and centimeter wavelengths \citep{Karnath2020} found that they were dominated by very bright, compact (R$\sim$100~au) emission, which sometimes had indications of complex morphological structure and at least one PBRS currently lacks evidence of an outflow. The compact structures could be a dense inner envelopes or very massive disks; measurements of their kinematics from molecular line emission have proven elusive thus far due to the high continuum opacity on scales $<$200~au and observations at centimeter wavelengths may be necessary to detect line emission \citep{DeSimone2020}.

\section{Summary and Outlook}

The advent of ALMA and space-based infrared observatories have truly revolutionized our understanding of protostellar systems. A decade and a half ago, it was largely unclear whether protostellar disks were common, and our picture of envelope structure had evolved minimally beyond the picture of a roughly spherical cow. Today, we have a clear picture of protostellar disks, including an increasingly detailed picture of their azimuthal and vertical structures. There is an emerging picture of streamers as a source of material flowing from envelope to disk, and a growing body of protostar masses that a decade ago might have seemed inconceivable. With that in mind, we finish by providing a number of points to summarize this review, along with a collection of issues that we expect will be tackled in the coming decade:

\begin{summary}[SUMMARY POINTS]
\begin{enumerate}
\item The envelopes around protostars are complex, with large scale asymmetric structure and a growing body of evidence for streamers that may be supplying material to their protostars and disks unevenly on a range of spatial scales.
\item There is strong evidence that all protostars have a rotationally-supported disk around them, and a distribution of initial angular momenta, coupled with magnetic fields likely play a role in shaping the broad distribution of disk sizes that are observed.
\item Disks around most protostars have larger submillimeter/millimeter luminosities than the disks around more-evolved pre-main sequence stars, but there remains substantial uncertainty in how these luminosities manifest from the total solid mass in disks.
\item While substructures are not absent in protostellar disks, there is early evidence that the significant features seen in protoplanetary disks only appear in earnest towards the end of the embedded phase.
\item There are $\sim$30 protostars with masses measured by a variety of techniques, but this distribution does not reflect the stellar initial mass function and there is no clear relationship between protostar mass and bolometric luminosity.

\end{enumerate}
\end{summary}

\begin{issues}[FUTURE ISSUES]
\begin{enumerate}
\item Diagnostics of protostellar evolution need to be reassessed in order to better relate observational characteristics to evolution with less influence from geometric orientation.
\item The impact of envelope structure on the forming disks needs to be assessed, in particular to better understand the frequency of streamers representing true overdensities in the envelope and thereby to what extent these streamers impact disk structure, evolution, and stability.
\item The translation of disk dust and gas emission to disk properties needs to be better characterized in order to under understand how fundamental disk properties, such as mass and radius in both gas and dust as well as the distribution of dust grain sizes, evolve across the star and planet formation process.
\item The gravitational stability of the protostellar disks needs to be better characterized to understand the role of gravitational instability in angular momentum transport, companion formation, and possibly giant planet formation.
\item An unbiased distribution of protostar masses should be measured to better understand the mass assembly timescale of protostars and the relation of the protostellar mass function to the luminosity function.
\item Further characterization of the candidates for the youngest protostars is needed to fully understand their relevance to the earliest stages of star formation.
\item The tools used to study the material around protostars need to grow in step with the growing complexity of questions being addressed so as to derive rigorous and quantitative conclusions in the face of the difficulty of disentangling structure from optical depth effects and the complicated nature of interferometric imaging.
\end{enumerate}
\end{issues}

\section*{DISCLOSURE STATEMENT}
The authors are not aware of any affiliations, memberships, funding, or financial holdings that
might be perceived as affecting the objectivity of this review. 

\section*{ACKNOWLEDGMENTS}
We are grateful to Leslie Looney, Ewine van Dishoeck, and Lynne Hillenbrand for insightful comments that helped improve this review. We wish to thank Jaime Pineda and Maria Teresa Valdivia-Mena for the availability of the code and data on streamers. We thank the ALMA eDisk team for the availability of their data for incorporation into the review figures. We thank Yancy Shirley and Brian Mason for the GBT Mustang 3~mm data of L1527 IRS. We thank Adele Plunkett for feedback on the protostar evolution cartoon. We thank Fabian Heitsch, Lee Hartmann, Ted Bergin, Merel van 't Hoff, Sarah Sadavoy, Chat Hull, Tyler Bourke, Mike Dunham, Satoshi Yamamoto, Nami Sakai, Yoko Oya, Amy Stutz, Daniel Harsono, Leslie Looney, Zhi-Yun Li, Daniel (Zhe-Yu) Lin, Jonathan Williams, Tom Megeath, Maria Jose Maureira, and Doug Johnstone for many insightful discussions over the years. This list is obviously incomplete and we hope that those not mentioned will forgive us! J.J.T acknowledges support from NASA XRP 80NSSC22K1159.
This paper makes use of ALMA data from numerous projects. ALMA is a partnership of ESO (representing its member states), NSF (USA) and NINS (Japan), together with NRC (Canada), MOST and ASIAA (Taiwan), and KASI (Republic of Korea), in cooperation with the Republic of Chile. The Joint ALMA Observatory is operated by ESO, AUI/NRAO and NAOJ.
The National Radio Astronomy Observatory is a facility of the National Science Foundation operated under cooperative agreement by Associated Universities, Inc.
\bibliographystyle{ar-style2}
\bibliography{references}

\begin{table}[ht]
\tabcolsep6.5pt
\caption{Protostar Masses}
\label{tab1}
\begin{center}
\begin{tabular}{@{}l|c|c|c|c|c|c|l@{}}
\hline
Source & Mass Upper & Mass Lower & Mass & T$_{\mathrm bol}$ & L$_{\mathrm bol}$ & Dist.  & Ref. \\
       &  (\msun)  & (\msun)    & (\msun) & (K)            & (L$_{\odot}$)     & (pc)    &           \\
\hline
HOPS-370   &  ...  &     ...  &   2.5  &   71   &   314 &   389  & \citep{Tobin2020b}\\
RCrA IRS7B & 3.21  &     2.09  &  2.65 &   88   &   5.1 &   152  & \citep{Ohashi2023}\\
L1527 IRS  & 0.49  &     0.32  & 0.41 &   41   &   1.3 &   140  & \citep{vanthoff2023}\\
L1489 IRS  & 1.91  &    1.5  & 1.7   &  213   &   3.4 &   146  & \citep{Yamato2023}\\
IRAS 04302+2247  & 1.65 & 1.23 &  1.44 &   88   &  0.43 &   160  & \citep{Lin2023}\\
CB68       & 0.158   &    0.137    & 0.15  &   50   &  0.89 &   151  & \citep{Kido2023}\\
Ced110 IRS4 & 1.45  &   1.21   & 1.33  &   68   &   1.0 &   189  & \citep{Sai2023}\\
IRAS 16253-2429   & 0.34 & 0.09  & 0.22     &   42   &  0.16 &   139  & \citep{Aso2023}\\ 
RCrA IRS5N & 0.4  &     0.18 &  0.29   &   59   &  1.4  &   147  & \citep{Sharma2023}\\
IRAS 15398-3359  &  ... & ... & 0.01    &   50   &  1.4  &   155  & \citep{Yen2017}\\ 
Oph IRS43-A  &  ... & ...     & 1.0    &  193   &  4.1  &   137  & \citep{Brinch2016}\\
Oph IRS43-B  &  ... & ...     & 1.0    &  193   &  4.1  &   137  & \citep{Brinch2016}\\
Oph IRS63  &  0.66 & 0.33     & 0.5    &  348   &  1.3  &   132  & \citep{Flores2023}\\
TMC1A      &  ...   & ...     & 0.64   &  183   &  2.3  &   137  &\citep{Aso2015}\\
B335       &  ... & ...       & 0.05   &   41   &  1.4  &   165  & \citep{Yen2015}\\ 
Lupus 3 MMS &   ... & ... &     0.3    &   39   &  0.41 &   200  & \citep{Yen2017}\\ 
VLA1623$^{\rm a}$    &   ... & ... &      0.2    &   50   &  1.1  &   140  & \citep{Murillo2013}\\
L1455 IRS 1 &  ... & ... &     0.28    &   59  &  3.6  &   300  &  \citep{Chou2016}\\
L1448IRS3B$^{\rm a}$  &  ... &    ...  & 1.19    &   61  &  5.8  &   300  & \citep{Reynolds2021}\\
L1448IRS3A  &  ... &    ...  & 1.51    &   47  &  5.8  &   300  & \citep{Reynolds2021}\\
TMC1$^{\rm a}$        &  ... &    ...  & 0.54    &   101  &  0.9  &   140  & \citep{Harsono2014}\\
L1551NE$^{\rm a}$     &  ... &    ...  &  0.8    &   91  &  4.2  &   140  & \citep{Takakuwa2012}\\
Elias29     &  ... &    ...  &  2.5    &   350  & 14.1  &   130  & \citep{Lommen2008}\\
HH212 MMS    &  ... &    ...  &  0.2    &  53 & 14    &   400  & \citep{Lee2017}\\
HH111 MMS    &  ... &    ...  &  1.8    &  78   & 17.4  &   400 & \citep{Lee2014}\\
L1551 IRS5$^{\rm a}$   &  ... &    ...  &  0.5    &  94   & 22.1  &   140 & \citep{Chou2014}\\
IRAS16293-2422-Aa & ... &  ... & 0.9   &  54  &  36   &   130 & \citep{Maureira2020}\\ 
IRAS16293-2422-Ab & ... & ...  & 0.8   &  54  &  36   &   130 & \citep{Maureira2020}\\ 
HOPS-361-A$^{\rm a}$        & 5.46 &  4.40 & 4.93   &  69  &  368   &   430 & \citep{Cheng2022}\\ 
HOPS-361-C$^{\rm a}$        & 1.57 & 1.38  & 1.48   &  69  &  85   &   130 & \citep{Cheng2022}\\ 

\end{tabular}
\end{center}
\begin{tabnote}
List of published mass measurements for known protostars. All the measurements are from molecular line kinematics except IRAS 16293-2422 and Oph IRS43 whose protostars have been observed over a long enough period of time to enable orbit fitting. The columns Mass Upper and Mass Lower refer to masses fit using the `Ridge' and `Edge' methods as described in Section 4.1; the Mass column is the average of the Upper and Lower masses if both are available.
\newline$^{\rm a}$Circumbinary mass measurement.
\end{tabnote}
\end{table}

\end{document}